\documentstyle[12pt]{article}
\input psfig.sty
\def\nn{\noindent}
\def\Re{{\cal R \mskip-4mu \lower.1ex \hbox{\it e}\,}}
\def\Im{{\cal I \mskip-5mu \lower.1ex \hbox{\it m}\,}}
\def\ie{{\it i.e.}}
\def\eg{{\it e.g.}}
\def\etc{{\it etc}}
\def\etal{{\it et al.}}

\def\sub#1{_{\lower.25ex\hbox{$\scriptstyle#1$}}}
\def\tev{\,{\ifmmode\mathrm {TeV}\else TeV\fi}}
\def\gev{\,{\ifmmode\mathrm {GeV}\else GeV\fi}}
\def\mev{\,{\ifmmode\mathrm {MeV}\else MeV\fi}}
\def\to{\rightarrow}

\def\subw{_{\rm w}}
\def\mh{\ifmmode m\sbl H \else $m\sbl H$\fi}
\def\mch{\ifmmode m_{H^\pm} \else $m_{H^\pm}$\fi}
\def\mt{\ifmmode m_t\else $m_t$\fi}
\def\mc{\ifmmode m_c\else $m_c$\fi}
\def\mz{\ifmmode M_Z\else $M_Z$\fi}
\def\mw{\ifmmode M_W\else $M_W$\fi}
\def\mws{\ifmmode M_W^2 \else $M_W^2$\fi}
\def\mhs{\ifmmode m_H^2 \else $m_H^2$\fi}   
\def\mzs{\ifmmode M_Z^2 \else $M_Z^2$\fi}
\def\mts{\ifmmode m_t^2 \else $m_t^2$\fi}
\def\mcs{\ifmmode m_c^2 \else $m_c^2$\fi}
\def\mchs{\ifmmode m_{H^\pm}^2 \else $m_{H^\pm}^2$\fi}
\def\ztwo{\ifmmode Z_2\else $Z_2$\fi}
\def\zone{\ifmmode Z_1\else $Z_1$\fi}
\def\mtwo{\ifmmode M_2\else $M_2$\fi}
\def\mone{\ifmmode M_1\else $M_1$\fi}
\def\tb{\ifmmode \tan\beta \else $\tan\beta$\fi}
\def\xw{\ifmmode x\subw\else $x\subw$\fi}
\def\ch{\ifmmode H^\pm \else $H^\pm$\fi}
\def\lum{\ifmmode {\cal L}\else ${\cal L}$\fi}
\def\inpb{\,{\ifmmode {\mathrm {pb}}^{-1}\else ${\mathrm {pb}}^{-1}$\fi}}
\def\infb{\,{\ifmmode {\mathrm {fb}}^{-1}\else ${\mathrm {fb}}^{-1}$\fi}}
\def\epem{\ifmmode e^+e^-\else $e^+e^-$\fi}
\def\ppb{\ifmmode \bar pp\else $\bar pp$\fi}
\def\bsg{\ifmmode B\to X_s\gamma\else $B\to X_s\gamma$\fi}
\def\bsll{\ifmmode B\to X_s\ell^+\ell^-\else $B\to X_s\ell^+\ell^-$\fi}
\def\bstt{\ifmmode B\to X_s\tau^+\tau^-\else $B\to X_s\tau^+\tau^-$\fi}
\def\lamt{\ifmmode \tilde\lambda\else $\tilde\lambda$\fi}
\def\shat{\ifmmode \hat s\else $\hat s$\fi}
\def\that{\ifmmode \hat t\else $\hat t$\fi}
\def\uhat{\ifmmode \hat u\else $\hat u$\fi}

\newskip\zatskip \zatskip=0pt plus0pt minus0pt
\def\matth{\mathsurround=0pt}

\def\atversim#1#2{\lower0.7ex\vbox{\baselineskip\zatskip\lineskip\zatskip
  \lineskiplimit 0pt\ialign{$\matth#1\hfil##\hfil$\crcr#2\crcr\sim\crcr}}}

\renewcommand{\thefootnote}{\fnsymbol{footnote}}

\begin{document} \begin{titlepage} 
\rightline{\vbox{\halign{&#\hfil\cr
&SLAC-PUB-8218\cr
&September 1999\cr}}}
\begin{center}

{\Large\bf Testing the Nature of Kaluza-Klein Excitations at Future Lepton 
Colliders}
\footnote{Work supported by the Department of 
Energy, Contract DE-AC03-76SF00515}
\medskip

\normalsize 
{\large Thomas G. Rizzo } \\
\vskip .3cm
Stanford Linear Accelerator Center \\
Stanford University \\
Stanford CA 94309, USA\\
\vskip .3cm

\end{center}

\begin{abstract} 
With one extra dimension, current high precision electroweak data constrain 
the masses of the first Kaluza-Klein excitations of the Standard Model gauge 
fields to lie above $\simeq 4$ TeV. States with  masses not much larger than 
this should be observable at the LHC. However, even for first excitation 
masses close to this lower bound, the second set of excitations will be too 
heavy to be produced thus eliminating the possibility of realizing the 
cleanest signature for KK scenarios. Previous studies of heavy 
$Z'$ and $W'$ production in this mass range at the LHC have demonstrated that 
very little information can be obtained about their couplings to the 
conventional fermions given the limited available statistics and imply that 
the LHC cannot distinguish an ordinary $Z'$ from the degenerate pair of the 
first KK excitations of the $\gamma$ and $Z$. In this paper we discuss the 
capability of lepton colliders with center of mass energies significantly 
below the excitation mass to resolve this ambiguity. In addition, we examine 
how direct measurements obtained on and near the top of the first excitation 
peak at lepton colliders can confirm these results. For more than one extra 
dimension we demonstrate that it is likely that the first KK excitation is too 
massive to be produced at the LHC.
\end{abstract} 




\renewcommand{\thefootnote}{\arabic{footnote}} \end{titlepage}


\section{Introduction} 

If Kaluza-Klein excitations of the Standard Model(SM) gauge fields exist then 
analyses of precision electroweak data indicate that the masses of the first 
excitation of the $W$, $Z$, $\gamma$ and $g$ must be greater than 
$\simeq 4$ TeV in the case of one extra dimension. For such 
heavy masses the second set of excitations will 
lie beyond the reach of the LHC even at several times design luminosity. 
In addition, the limited statistics at such large invariant masses will 
($i$) most likely render the gluon excitation invisible due to both its large 
width to mass ratio as well as detector jet energy smearing and ($ii$) will not 
allow the photon and Z resonances to be resolved even if they are not exactly  
degenerate. Thus the LHC will see what appears to be a degenerate $Z'$ and 
$W'$, something that occurs in many more ordinary extended electroweak 
gauge models. Based upon 
past studies of new gauge boson coupling determinations at hadron colliders 
we know that with the available statistics the LHC will not be able to identify 
these resonances as KK excitations.
How can we resolve this issue? As we will show below, a lepton collider, even 
one operating reasonably far below the apparent $Z'$ resonance will most 
likely provide evidence compelling enough to resolve this 
ambiguity. Furthermore, 
we will demonstrate that a higher energy lepton collider, sitting on this 
resonance peak, will very easily distinguish the two possibilities not via an 
analysis of the line shape but through several 
factorization tests among electroweak observables. The extension to the case 
of more than one extra dimension is also discussed. 

This paper is organized as follows: Section 2 contains a pedagogical theory 
background for the definition of the problem and the analysis that follows. 
The details of the 
problem outlined above are discussed in Section 3. In Section 4 we discuss 
how a linear collider operating at energies well below the mass of the first 
KK excitations, presumed discovered at the LHC, will yield strong evidence 
about its 
fundamental nature while in Section 5 we discuss what additional information 
can be learned by sitting on the KK resonance at a future lepton collider. A 
summary and further discussion is given in Section 6.

\section{Background}

String/M-theory tells us that we live in a world with at least 
six extra dimensions. 
It is perhaps likely that the size of these dimensions are of order the 
inverse Planck scale, $\sim 1/M_{pl}$, and may remain forever hidden from 
direct experimental confirmation. However, in the past two years the 
possibility has re-emerged{\cite {old}} that at least some of 
these extra dimensions may be 
much larger and not far away from the electroweak scale, $\sim 1/$TeV, that 
is now being 
probed at colliders. In one appealing scenario{\cite {nima}} due to 
Arkani-Hamed, Dimopoulos and Dvali(ADD), gravity is 
allowed to propagate in at least two `large' extra dimensions while the fields 
of the SM are confined to D-branes of appropriate dimension 
transverse to these. (Here by `large' we mean compactification radii 
$>>1/$TeV.) Such a structure 
allows for the Planck scale to be brought down from $10^{19}$ GeV to only a 
few TeV offering a new slant on the hierarchy problem. The specific size of 
these `large' dimensions depend on how many we assume there to be; for $n$ 
extra `large' dimensions the common compactification radius $R$ is order 
$\sim 10^{30/n-19}$ m. The rich phenomenology of this model has been examined 
in a very rapidly growing series of papers{\cite {pheno}}. (We note in 
passing that the ADD 
scenario assumes that the metric tensor on the brane does not depend on the 
compactified co-ordinates, \ie, that it factorizes; this need not be 
necessary{\cite {ran}}.) If $n<6$ 
there can also be some extra `small' longitudinal dimensions wherein both 
the SM fields as well as gravity can live. (Here by `small' we mean 
compactification radii $\sim 1/$TeV.) For example, we 
can imagine there being 4 `large' extra dimensions in which only gravity 
propagates and 2 `small' extra dimensions populated by both gravity as well 
as the SM gauge fields. The propagation 
of the SM fields into these `small' dimensions can lead to a drastic lowering 
of the GUT scale{\cite {guts}} due to an almost power-like running of the 
couplings. There are many variations on this particular theme depending upon 
which and how many SM fields we allow to feel the extra dimensions. In what 
may be the most well motivated and 
attractive scheme, and the one we consider below, only the SM 
gauge fields (and the Higgs field) can propagate in the extra 
dimensions while the chiral fermions only experience the usual four dimensions 
and thus lie on a 3-brane, \ie, `the wall'. (We imagine that all of the SM 
gauge fields feel the same number of the extra dimensions in what follows.) 
It is now possible to imagine a viable scenario wherein the Planck, 
string, compactification and GUT scales are not too far above a few TeV. 

In addition to probing weak scale gravity{\cite {pheno}} another test of this 
scenario is to search for the Kaluza-Klein(KK) excitations of the SM gauge 
fields. In fact, the hallmark{\cite {old,bumps}} of these KK theories is the 
existence of regularly spaced 
resonances in the $\ell^+\ell^-$, $\ell^{\pm} \nu$ and $jj$ channels at hadron 
colliders, such as the Tevatron and LHC, which are degenerate, \ie, 
the first excitations, $\gamma^{(1)},Z^{(1)},W^{(1)}$ and $g^{(1)}$, 
have a common mass, in the limit that 
mixing with the corresponding SM zero modes is neglected. (In practice, even 
when mixing is present the fractional mass shifts are quite negligible 
for KK states above 1 TeV.) As we will see below, such recurrence structures 
are not always observable making the direct experimental case for KK 
scenarios less transparent. For one extra longitudinal dimension, 
compactification on $S^1/Z_2$ leads to equally spaced states with masses 
given by $M_n=n/R$ and with a 
non-degenerate level structure. Due to the normalization of the gauge field 
kinetic energies, the excitations in the KK tower naively couple to the SM 
fermions with a strength larger than that of the zero-modes by a universal 
factor of $\sqrt 2$ assuming that the fermions are all localized 
at the same fixed point on the wall{\cite {schm}}. (More on this point below.) 
For the case of 
more than one extra dimension the situation is far more complex and 
depends upon the details of the compactifying manifold. Here we find that 
not only are the KK excitation spacings more intricate but many of the levels 
become 
degenerate and the strength of the coupling in comparison to the zero-modes 
becomes level dependent. For example, in 
the case of two additional dimensions with a 
$S^1/Z_2 \times S^1/Z_2$ compactification, assuming both compactification 
radii are equal, the first five KK levels 
occur at masses of (in units of $1/R$) $1, \sqrt 2, 2, \sqrt 5$ and $\sqrt 8$ 
with degeneracies of $2,1,2,2$ and 1 and with 
naive relative coupling strengths of 
$\sqrt 2,2,\sqrt 2,2$ and 2, respectively. Alternative compactifications 
yield other more intricate patterns as do extensions to the case of 
even more dimensions.

What do we know about the size of the compactification radii for these 
longitudinal dimensions, \ie, what bounds are there on the masses of the SM 
excitations? From direct searches for $Z'$, $W'$ and dijet bumps 
at the Tevatron{\cite {tev}}, it is clear 
that the masses of the first tower states are in excess of $\sim 0.7-0.9$ TeV. 
Through cross section and asymmetry measurements at LEP II the anticipated 
reach for the first 
KK state through {\it indirect} means will be approximately 3 TeV by combining 
the results of all four experiments assuming adequate luminosity is achieved. 
However, by examining the influence of KK towers on electroweak 
measurements{\cite {fits}} we can 
place far tighter bounds on $1/R$, or equivalently, 
the first excitation masses {\it provided} we make 
a number of assumptions. First, as is usual in these types of analyses, it is 
assumed that the KK fields are the {\it only} source of new physics that 
perturb the SM predictions for electroweak quantities. Secondly, we must 
assume that the couplings of at least the first few recurrences to the 
SM fields are 
not vastly different than those given by the simple rescaling correction 
due to the normalization of the gauge field kinetic energies  
discussed above. The reason to worry about this particular assumption is 
clear by considering the 
limit wherein the effects of KK tower exchanges can be written as a set of 
contact 
interactions by integrating out the tower fields. Almost all of the current 
constraints on the masses of KK states arise from consideration of this 
contact interaction limit. In this case tower exchanges 
lead to new dimension-six operators whose coefficients can be shown to be 
proportional to a fixed dimensionless quantity, $V$, which can be symbolically 
written as{\cite {us}} 
\begin{equation}
V=(M_wR)^2\sum_{{\bf n}=1}^\infty {g_{\bf n}^2\over {g_0^2}} {1
\over {{\bf n} \cdot {\bf n}}} \,,
\end{equation}
where $g_{\bf n}$ is the coupling of the 
$n^{th}$ KK level labelled by the set of integers {\bf n}. $M_w$, 
is  the $W$ boson mass which we employ as a typical weak scale. 
(Here for simplicity of presentation we have assumed a 
$Z_2\times Z_2\times...$ compactification so that the first KK excitation(s) 
has a mass $1/R$.) Through an analysis of precision measurements the value 
of $V$ can be directly restricted thus leading to an apparent constraint on 
$R$ for any given number of extra dimensions and specified 
compactification scenario. However, this seemingly 
straightforward program runs into an immediate difficulty requiring 
a somewhat detailed digression. The resolution of this difficulty has 
influence upon where we anticipate the mass of the KK excitations to lie and, 
through possible modifications their couplings to the fermions of the SM, 
their production cross sections at colliders. 

Using the naive scaling of the 
couplings the sum in the expression above only converges (to a value 
of $2\sum_{n=1}^\infty 1/n^2=\pi^2/3\simeq 3.28987$ assuming all 
the fermions are properly localized) in the case 
of a single extra dimension. There are several 
ways to deal with this problem. The first and most often used{\cite {guts}} 
approach is to sum over a finite number of terms, \ie, only 
those states whose masses lie below the string scale, $M_s$, which 
now acts simply as a cut off. For example, in the case of one extra 
dimension, we cut off the sum at $n=n_{max}\simeq M_sR$ and for any fixed 
assumed value of $M_sR$ we will of course obtain a smaller value than given by 
the complete sum. If 
$n_{max}=5(10,20)$ we obtain for the partial sum 
2.92722(3.09954,3.19233), which are all not far from the value above
due to the rapid convergence of the series. While this procedure does not 
numerically reduce the sum in any serious manner 
in the one dimensional case it has a far greater 
influence in more than one dimension since this partial
 sum is finite. For example, 
in the case of $Z_2\times Z_2$ taking $n_{max}=5$, so that we include 
only the first 14 mass states in the tower, yields a value of 12.7826
for the partial sum. Note that this is appreciably larger than in the 
one dimensional case. Taking instead 
$n_{max}=10(20)$ yields the corresponding results of 17.0790(21.4083) which 
shows an approximate logarithmic growth with increasing $n_{max}$. 
Given a fixed 
upper bound on the value of $V$ this would imply that the lower bound on the 
mass of the first KK excitation would have to be at least a factor of 2 larger 
than in the one-dimensional case. Even though the cross section for the 
production of this state would be 4 times larger than in the one dimensional 
case due to the enhanced coupling it is clear that 
this state would be to massive to be produced at the LHC.
While this approach regularizes the tower sum, this straightforward 
truncation technique appears to be somewhat arbitrary and 
conceptually inadequate.

A second possibility is that the KK couplings to four dimensional fields have 
an {\it additional} level 
dependence, above and beyond that due to the appropriate kinetic energy 
normalization factor, that exponentially damps the contributions from higher 
terms in the sum{\cite {wow}}. Such a suppression has been suggested on several 
grounds including the high energy behaviour of string scattering 
amplitudes and also through 
considerations of the flexibility of the wall{\cite {wow}}. In the later 
case, for a rather rigid wall, the still infinite sum is now of the form
\begin{equation}
V=(M_wR)^2\sum_{{\bf n}=1}^\infty {g_{\bf n}^2\over {g_0^2}} 
{1\over {{\bf n} \cdot {\bf n}}} e^{-{\bf n} 
\cdot {\bf n}/n_{max}^2} \,,
\end{equation}
where $n_{max}$ is as given above. For any given value of $n_{max}$ the 
sum is finite and, if $n_{max}$ is not too small the couplings of the first 
KK excitations hardly differ from their naive values. We note that for the 
case of one extra dimension taking $n_{max}=5(10,20)$ yields the sum 
2.62089(2.94538, 3.11512), again not far from actual naive sum but somewhat 
smaller than in the simpler truncation of the summation approach. 
In addition, by absorbing the exponential into the definition of their 
strengths, the 
corresponding couplings of the first excitation relative to the zero mode is 
given by $g_1^2/2g_0^2=0.9608(0.9901,0.9975)$, respectively. Here we see that 
there is very little suppression in the strength of the couplings 
in comparison to the naive value obtained through from the gauge field kinetic 
terms. (We note 
that if we assume that the wall is {\it not} rigid then all terms in the sum 
become very highly suppressed and the KK excitations almost completely 
decouple from the SM fermions. In this case, unfortunately, nothing can be 
said about the KK excitation masses from experimental data.) If we extend 
this same approach to the case of two extra dimensions with a $Z_2\times Z_2$ 
compactification where the sum is 
naively divergent, assuming as above that $n_{max}=5(10,20)$, 
we now obtain a result of 
6.73478(8.91208, 11.08966). Here we see that the exponential cut off 
approach is actually more efficient than the ad hoc termination of the series 
for fixed $n_{max}$. 
As we will see below, if $n_{max}=5$ is realized, then the 
lower bound on the 
mass of the first KK excitation in this two-dimensional scenario 
is $\simeq 5.6$ TeV which should be visible at the LHC due to the 
coherence in the production cross section among the degenerate states. 
However, we note the important result that for larger values of $n_{max}$, 
the bound from $V$ on the masses of the first KK excitations will drive 
them {\it beyond} the reach of the LHC. Clearly, this result persists and 
becomes even stronger when we 
go to the case of more than two extra dimensions.

\section{The Problem}

Given these digressions, the analysis of the precision electroweak data as 
presented at Moriond{\cite {mor}} by the authors of Ref.{\cite {us}} yielded 
the constraints $V \leq 0.0015-0.0020$ depending upon what fraction of the SM 
Higgs vacuum expectation value arises from a Higgs in the bulk. With the 
improved data presented at the summer conferences{\cite {mnich,lp99}}, we 
would expect these bounds to slightly tighten. Repeating the analysis as 
presented in Ref.{\cite {us}} with this new data, and assuming that the Higgs 
boson mass is $\geq 100$ GeV{\cite {higgs}}, yields a somewhat stronger 
bound of 
$V \leq 0.0010-0.0013$. Given the discussion above 
it is reasonably straightforward to interpret these results in the case of one 
extra dimension for reasonable values of $n_{max}$ using either approach: we 
obtain 
$1/R=M_1\geq 3.9$ TeV, where $M_1$ is the mass of the first KK excitation. 
(A similar, somewhat weaker, but more model independent bound of $M_1>3.4$ TeV 
can be obtained from existing constraints on charged current 
contact interactions as has been 
recently shown by Cornet, Relano and Rico in Ref. {\cite {fits}}.) In 
the case of two or more extra dimensions the bound is somewhat harder to 
interpret but it is clear from the above discussion and 
a short numerical study that the masses of the 
first KK excitations must be significantly larger than in the case of one extra 
dimension since the sum over states yields a significantly larger value. This 
result is very important in that it tells us that if the KK scenario is 
correct then ($i$) in the case of one extra dimension the radius of 
compactification and, hence, the masses of the first excitations must be such 
that the masses of the {\it second set} of excitations must lie above the 
reach of the 
LHC{\cite {bumps,us}} in both the Drell-Yan and $jj$ channels. This implies 
that the most obvious signal for the KK scenario will not be realized at the 
LHC even if KK excitations do exist. Also, ($ii$) as mentioned above 
for the case of two extra 
dimensions, even the masses of the {\it first} KK states will be beyond the 
reach of the LHC unless $n_{max}$ is quite small $\leq 5$. 
Table 1 summarizes our 
results for the lower bound on $M_1$ for different $n_{max}$ for various 
compactification scenarios in different dimensions employing either of the 
above cut off schemes. We see that even with a very small $n_{max}$ the value 
of $M_1$ is beyond the reach of the LHC in the case of three extra dimensions.

\vspace*{0.3cm}
\begin{table*}[htpb]
\leavevmode
\begin{center}
\begin{tabular}{lcccccc}
\hline
\hline
     &\multicolumn{2}{c}{ $Z_2\times Z_2$}  &\multicolumn{2}{c}{$Z_{3,6}$} & 
\multicolumn{2}{c}{$Z_2\times Z_2\times Z_2$} \\ 
\hline
\hline
$n_{max}$ &  T   &  E   &  T   &  E   &  T   &  E    \\
\hline
2   & 5.69$^*$  & 4.23$^*$  &6.63$^*$  &4.77$^*$ &8.65  & 8.01  \\  
3   & 6.64   & 4.87$^*$ & 7.41 &5.43$^*$ & 11.7 & 10.8 \\
4   & 7.20   & 5.28$^*$ & 7.95 &5.85$^*$ & 13.7 & 13.0 \\
5   & 7.69   & 5.58$^*$ & 8.36 &6.17$^*$ & 15.7 & 14.9 \\
10  & 8.89   & 6.42     & 9.61 &7.05  & 23.2  &  22.0  \\
20  & 9.95   & 7.16     & 10.2 &7.83  & 33.5  &  31.8  \\
50  & 11.2   & 8.04     & 12.1 &8.75  & 53.5  &  50.9  \\
\hline
\hline
\end{tabular}
\caption{Lower bound on the mass of the first KK state in TeV resulting from 
the constraint on $V$ for the case of more than one dimension. `T'[`E'] labels 
the result 
obtained from the direct truncation (exponential suppression) approach as 
discussed in the text. Cases 
labeled by an asterisk will be observable at the LHC. $Z_2\times Z_2$ and 
$Z_{3,6}$ correspond to compactifications in the case of two extra dimensions 
while $Z_2\times Z_2\times Z_2$ is for the case of three extra dimensions.}
\end{center}
\end{table*}
\vspace*{0.4cm}

\vspace*{-0.8cm}
\nn
\begin{figure}[htbp]
\centerline{
\psfig{figure=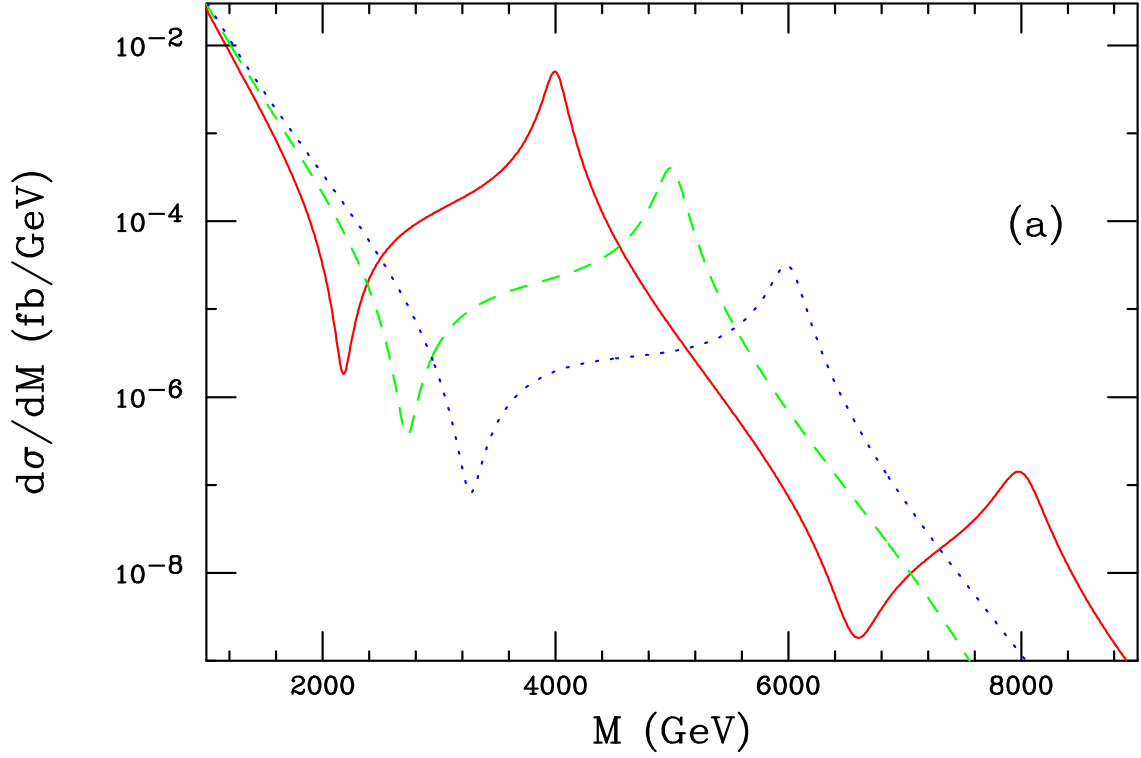,height=10.0cm,width=15cm,angle=90}}
\vspace*{9mm}
\centerline{
\psfig{figure=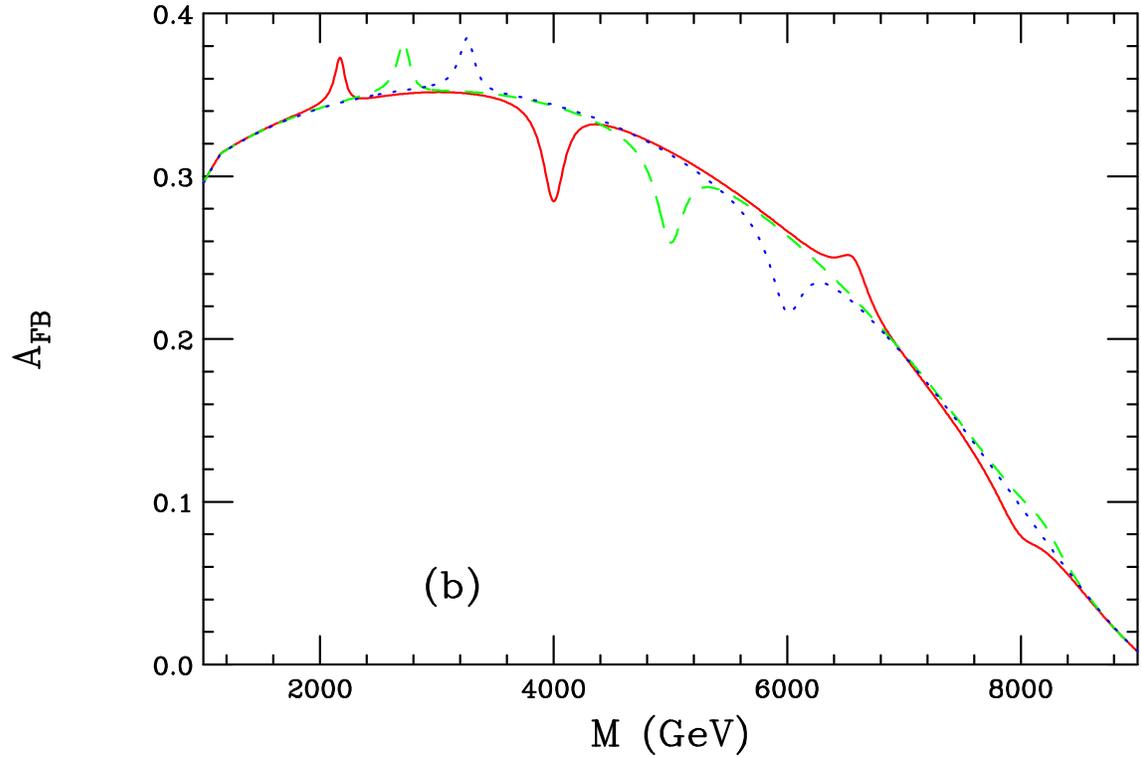,height=10.0cm,width=15cm,angle=90}}
\vspace*{0.0cm}
\caption{(a)Cross section and (b) Forward-Backward asymmetry for Drell-Yan 
production of the degenerate neutral KK excitations $Z^{(n)}$ and 
$\gamma^{(n)}$ as a function of the dilepton invariant mass at the LHC 
assuming one extra dimension and naive coupling values with $1/R$=4(5, 6) TeV 
corresponding to the solid(dashed, dotted) curve. The second excitation is 
only shown for the case of $1/R=4$ TeV.}
\end{figure}
\vspace*{0.4mm}
\vspace*{-0.8cm}
\nn
\begin{figure}[htbp]
\centerline{
\psfig{figure=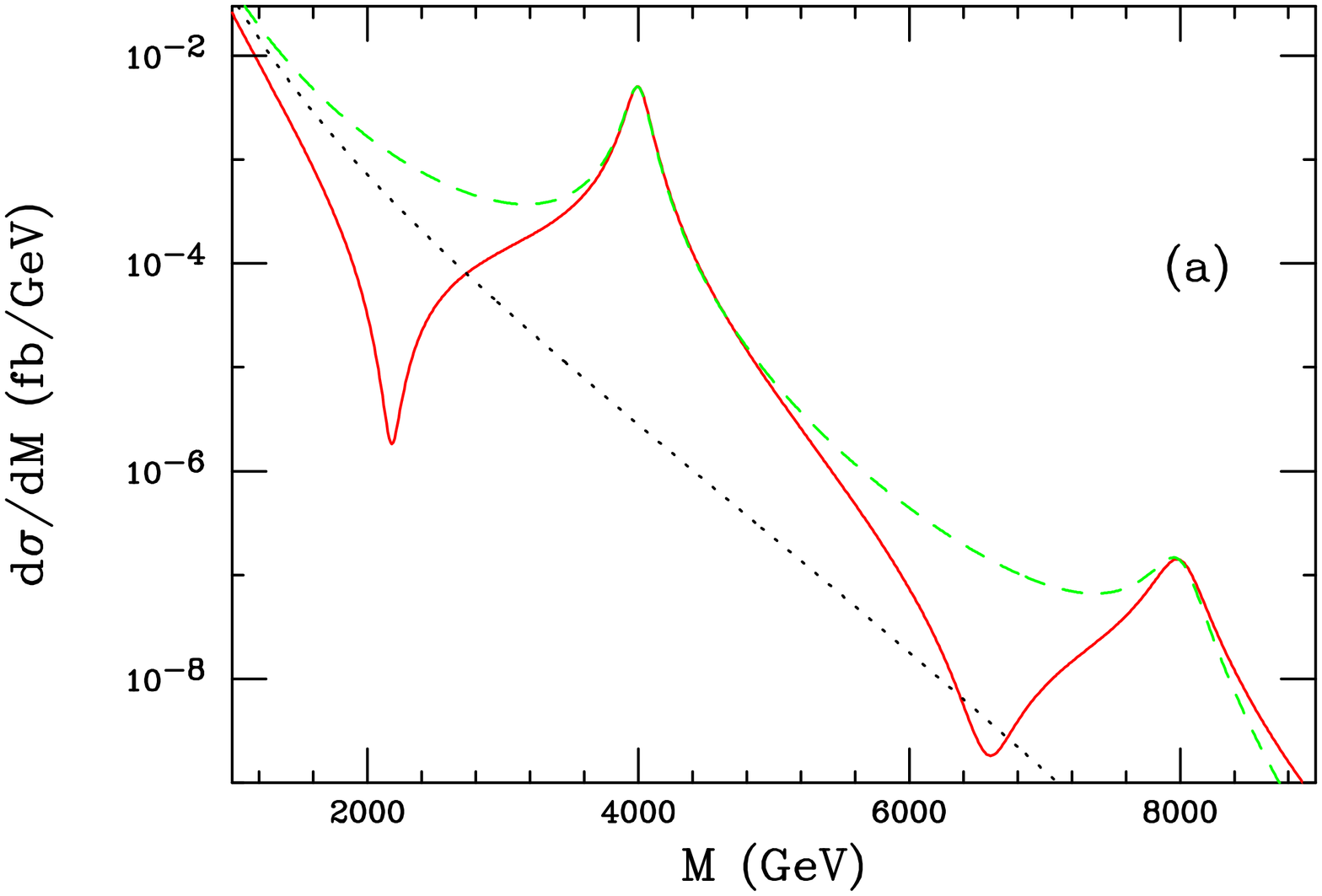,height=10.0cm,width=15cm,angle=90}}
\vspace*{9mm}
\centerline{
\psfig{figure=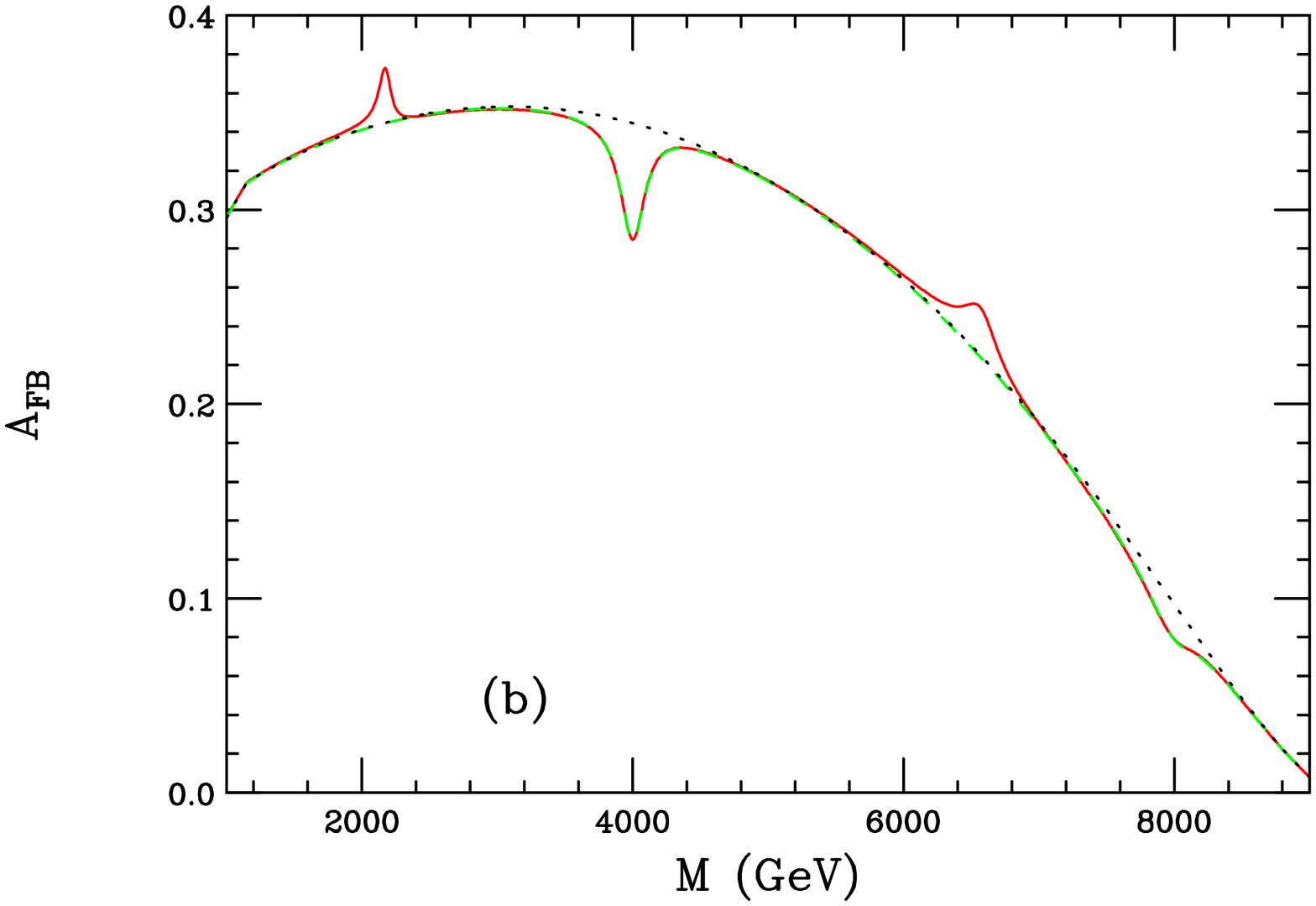,height=10.0cm,width=15cm,angle=90}}
\vspace*{0.0cm}
\caption{Same as the previous plot for the case of $1/R=4$ TeV(solid) but now 
also showing both the SM result for comparison(dots) as well as the alternative 
excitation curve for the case where the naive couplings are altered as in the 
scenario of Arkani-Hamed and Schmaltz(dashed){\cite {schm}}.}
\end{figure}
\vspace*{0.4mm}

At the LHC in the case of one extra dimension (or two extra dimensions with 
$n_{max}$ very small) we are left with the somewhat more subtle 
KK signature of degenerate $\gamma^{(1)}$, $Z^{(1)}$, $W^{(1)}$ and 
$g^{(1)}$ states. However, even this signature 
may not be realized experimentally as the $g^{(1)}$ resonance may be easily 
washed out due to both experimental jet energy resolution and the resonance's 
very large width to mass ratio{\cite {bumps}} when its 
mass lies in the range above $\simeq 4$ TeV. More than likely a shoulder-like 
structure would remain visible but would be difficult to 
interpret{\cite {nath}}. We are thus left with a degenerate set of 
$\gamma^{(1)}$, $Z^{(1)}$, and $W^{(1)}$ as a potential signal for 
KK scenarios at the LHC. Unfortunately a degenerate pair of new gauge bosons 
$Z',W'$ is {\it not} a unique signal for KK models as many extended electroweak 
theories predict{\cite {models}} such a situation. Of course the single 
`resonance' in 
the $\ell^+\ell^-$ channel in the KK case is actually a superposition of both 
the $\gamma^{(1)}$ and $Z^{(1)}$ and not just a $Z'$. Our claim here is that 
the LHC will not be able to distinguish these two possibilities given the 
rather small number of available observables 
due to the rather limited statistics. 

To clarify this situation let us consider the results displayed in 
Figs. 1 and 2 for the case of one extra dimension. In Fig.1 we show the 
production cross sections and Forward-Backward Asymmetries, $A_{FB}$, in the 
$\ell^+\ell^-$ channel with inverse 
compactification radii of 4, 5 and 6 TeV. In calculating these cross sections 
we have assumed that the KK excitations have their naive couplings and 
can only decay to the usual fermions of 
the SM. Additional decay modes can lead to appreciably lower cross sections so 
that we cannot use the peak heights to determine the degeneracy of the KK 
state. Note that in the 4 TeV case, which is essentially as small a mass 
as can be tolerated 
by the present data on precision measurements, the second KK 
excitation is visible in the plot. We see several things from these figures. 
First, we can easily estimate the total number of events in 
the resonance regions associated with each of the peaks assuming the canonical 
integrated luminosity of $100 fb^{-1}$ appropriate for the LHC; we find 
$\simeq 300(32, 3, 0.02)$ events corresponding to the 4(5,6,8) TeV 
resonances if we sum over both electron and muon final states and assume 
$100\%$ leptonic 
identification efficiencies. Clearly the 6 and 8 TeV resonances will 
not be visible at the LHC (though a modest increase of luminosity will allow 
the 6 TeV resonance to become visible) and we also verify our claim that 
only the first KK 
excitations will be observable. In the case of the 4 TeV resonance there is 
sufficient statistics that the KK mass will be well measured and 
one can also imagine measuring $A_{FB}$ since the 
final state muon charges can be signed. Given sufficient statistics, a 
measurement of the angular distribution would demonstrate that the state 
is indeed spin-1 and not spin-0 or spin-2. However, for such a heavy resonance 
it is unlikely that much further information could be obtained about its 
couplings and other properties and the values of $A_{FB}$ alone cannot 
determine whether this resonance is composite no matter how much statistics 
is available. In fact the conclusion of $Z'$ 
analyses{\cite {snow}} 
is that coupling information will be essentially impossible to obtain for 
$Z'$-like resonances with masses in excess of 1-2 TeV at the LHC. 
Furthermore, the lineshape of the 4 TeV resonance will be difficult to measure 
in detail due to both the limited statistics and energy smearing. Thus we 
will never know from LHC data alone whether the first KK resonance has been 
discovered or, instead, some extended gauge model scenario has been realized. 

It is often stated{\cite {bumps}} that the sharp dip in the cross section 
at an approximate dilepton pair invariant mass of $\simeq M_1/2$ 
will be a unique signal for the KK scenario. However, there 
are several difficulties with this claim. First, it is easy to construct 
alternative models with one extra dimension wherein either the leptonic or 
hadronic couplings of the odd excitations have opposite sign to the usual 
assignment, as in the model of Arkani-Hamed and Schmaltz(AS){\cite {schm}}, 
since 
the fermions lie at different fixed points on the wall. Here we consider the 
specific scenario where the quarks and leptons are at opposite fixed points. In 
this case the excitation curves will look quite 
different as shown in Fig.2 where we see that the dip below the resonance has 
now essentially disappeared. Second, even if the dip is present it will be 
difficult to observe directly 
given the LHC integrated luminosity. The reason here is 
that if we examine a 100 GeV wide bin around the location of the apparent 
minimum, the SM predicts only 5 events to lie in this bin 
assuming an integrated luminosity of 100 $fb^{-1}$. 
To prove that any dip is present we would need to demonstrate that the event 
rate is significantly below the SM value which would appear to be rather 
difficult if at all possible. Thus we stick to our conclusion that the 
LHC cannot distinguish between an ordinary $Z'$ and the degenerate 
$Z^{(1)}/\gamma^{(1)}$ resonance and we must turn elsewhere to resolve this 
ambiguity.

\section{Lepton Colliders Below the Resonance}

It is well-known that future $e^+e^-$ linear colliders(LC) operating in the 
center of mass energy range $\sqrt s=0.5-1.5$ TeV will be sensitive to indirect 
effects arising from the exchange of new $Z'$ bosons with masses typically 6-7 
times greater than $\sqrt s${\cite {snow}}. Furthermore, analyses have shown 
that with enough statistics the couplings of the new $Z'$ to the SM fermions 
can be extracted{\cite {coupl}} in a rather precise manner, especially when 
the $Z'$ mass is already approximately known from elsewhere, \eg, the LHC. (If 
the $Z'$ mass is not known 
then measurements at several distinct values of $\sqrt s$ can be used to 
extract both the mass as well as the corresponding couplings{\cite {me}}.) 

\vspace*{-0.5cm}
\nn
\begin{figure}[htbp]
\centerline{
\psfig{figure=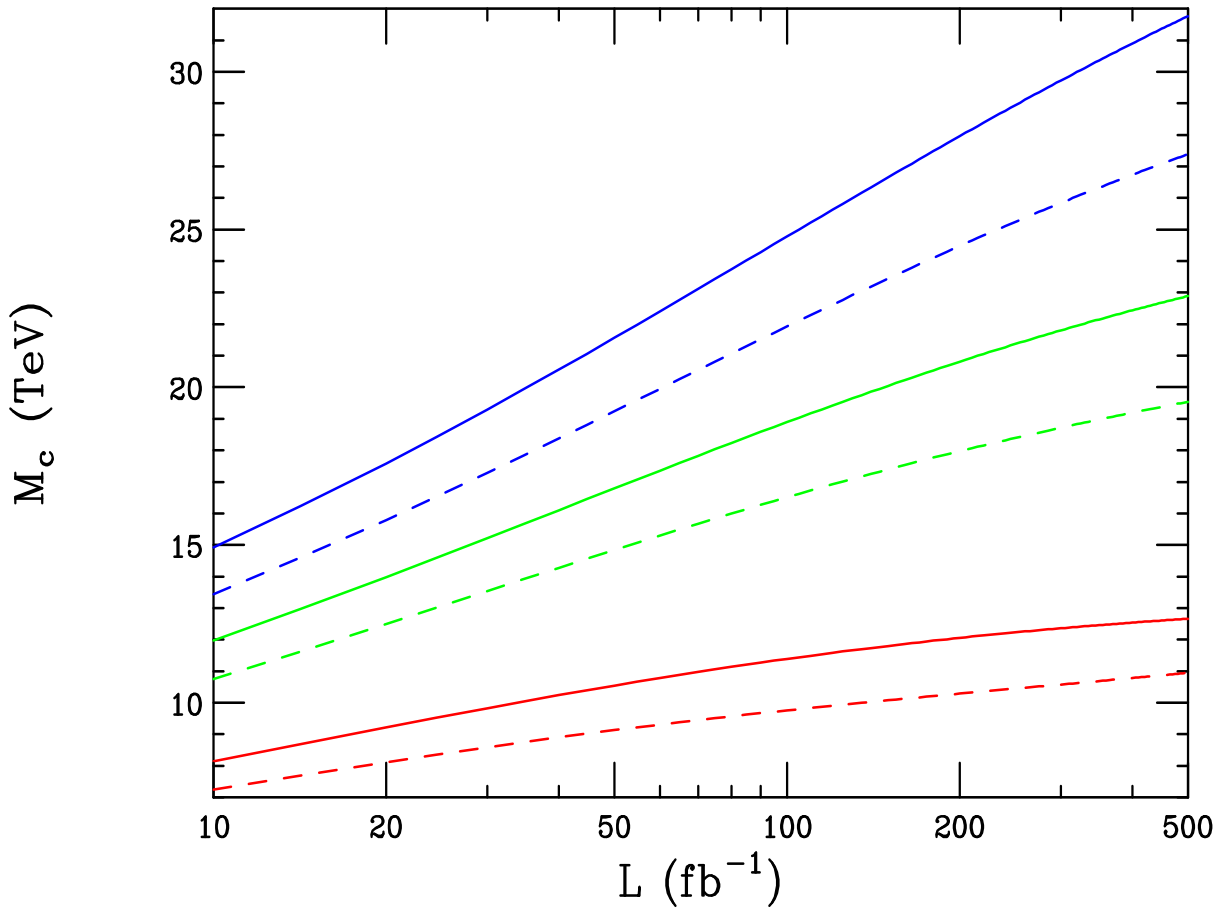,height=12.0cm,width=16cm,angle=90}}
\vspace*{0.1cm}
\caption[*]{Search reach, $M_c$, for the first $Z^{(1)}/\gamma^{(1)}$ excited 
state as a function of the integrated luminosity assuming $e^+e^-$ collider 
center of mass energies, from top to bottom, of 1.5, 1, 0.5 TeV. One extra 
dimension is assumed. The solid curves correspond to the case of `conventional' 
couplings while the dashed curves are for the case of the AS 
scenario{\cite {schm}}.}
\end{figure}
\vspace*{0.4mm}

In the present situation, we imagine that the LHC has 
discovered and determined the mass of a $Z'$-like resonance in the 4-6 TeV 
range. Can the LC tell us anything about this object? The first question to
ask is whether the LC can indirectly 
detect this excitation via the $e^+e^-\to f\bar f$ 
channels. (More precisely, can it probe the entire tower of KK states of which 
the 4-6 TeV object is the lowest lying one.)  To address this issue we have 
repeated the Monte Carlo analyses in {\cite {snow,coupl,me}} and have asked 
for the search reach for the first KK excitation as a function of integrated 
luminosity. To obtain our results we have combined the $f=e,\mu,\tau,b,c$ and 
$t$ final states, assumed $90\%$ beam polarization and included angle cuts, 
initial state radiation, identification efficiencies and systematics 
associated with the overall luminosity determination. The angular distribution 
of the various cross sections, the Left-Right Asymmetries, $A_{LR}^f$, and 
the polarization of the $\tau$'s in the final state are simultaneously 
combined in this fit. The search reaches are 
shown in Fig.3 for the case of one extra dimension and assume the 
conventional naive coupling relationships. Note that the reach is as much as 
three times greater than that for a more conventional $Z'$. The reasons for 
this are as follows: ($i$) the couplings of the KK excitations are larger 
than those of their SM partners by $\sqrt 2$, ($ii$) the complete towers 
contribute to these deviations and ($iii$) both 
$\gamma^{(n)}$ and $Z^{(n)}$ towers are present and add coherently. If we 
allow for more than one extra dimension, cutting off the KK sum by either of 
the techniques described above, it is clear that our resulting reach will 
be significantly higher in mass 
due to the greater number of states and the larger couplings involved.

\vspace*{-0.5cm}
\nn
\begin{figure}[htbp]
\centerline{
\psfig{figure=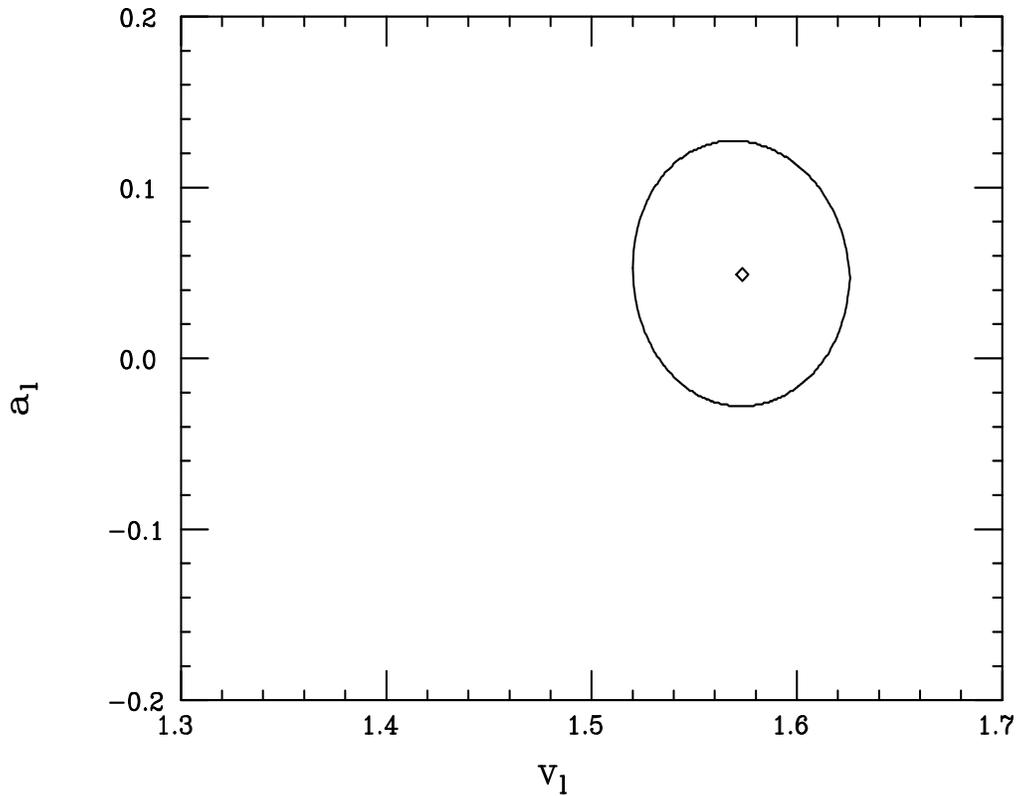,height=13cm,width=16cm,angle=-90}}
\vspace*{0.1cm}
\caption[*]{Fitted values of the parameters $v_l$ and $a_l$ following the 
procedures described in the text for a 4 TeV KK excitation at a 500 GeV  
$e^+e^-$ collider. The contour described the $95\%$ CL region with the best 
fit value as a diamond. The normalization is such that the corresponding SM 
$Z$ boson's axial-vector coupling to the electron is -1/2.}
\end{figure}
\vspace*{0.4mm}

The next step would be to use the LC to extract the couplings of the 
apparent resonance discovered by the LHC; we find that it is sufficient for 
our arguments to do this solely for the leptonic channels. The idea is the 
following: we measure the deviations in the differential cross sections 
and angular dependent $A_{LR}^f$'s 
for the three lepton generations and combine those with $\tau$ polarization 
data. Assuming lepton universality(which would be observed in the LHC data 
anyway), that the resonance mass is well 
determined, and that the resonance is an ordinary $Z'$ we perform a fit to 
the hypothetical $Z'$ coupling to leptons, $v_l,a_l$. To be specific, let us 
consider the case of only one extra dimension with a 
4 TeV KK excitation and employ a $\sqrt s=500$ GeV 
collider with an integrated luminosity of 200 $fb^{-1}$. The result of 
performing these fits is shown 
in Fig.4 from which we see that the coupling values are `well determined'
(\ie, the size of the allowed region we find is quite small) by the fitting 
procedure as we would have expected from previous analyses of $Z'$ couplings
extractions at linear 
colliders{\cite {snow,coupl,me}}. We note that identical results are 
obtained for this analysis if we assume that the KK excitations are of the 
type discussed by Arkani-Hamed and Schmaltz{\cite {schm}}.

\vspace*{-0.8cm}
\nn
\begin{figure}[htbp]
\centerline{
\psfig{figure=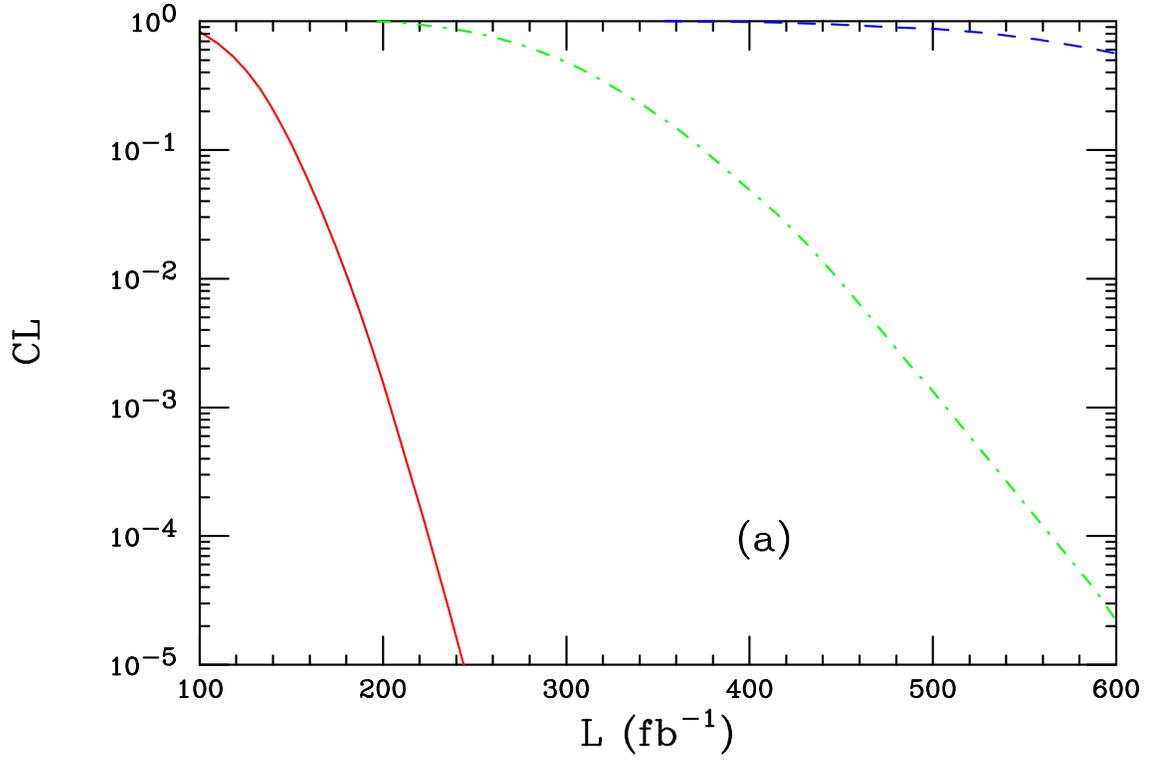,height=10.0cm,width=15cm,angle=90}}
\vspace*{9mm}
\centerline{
\psfig{figure=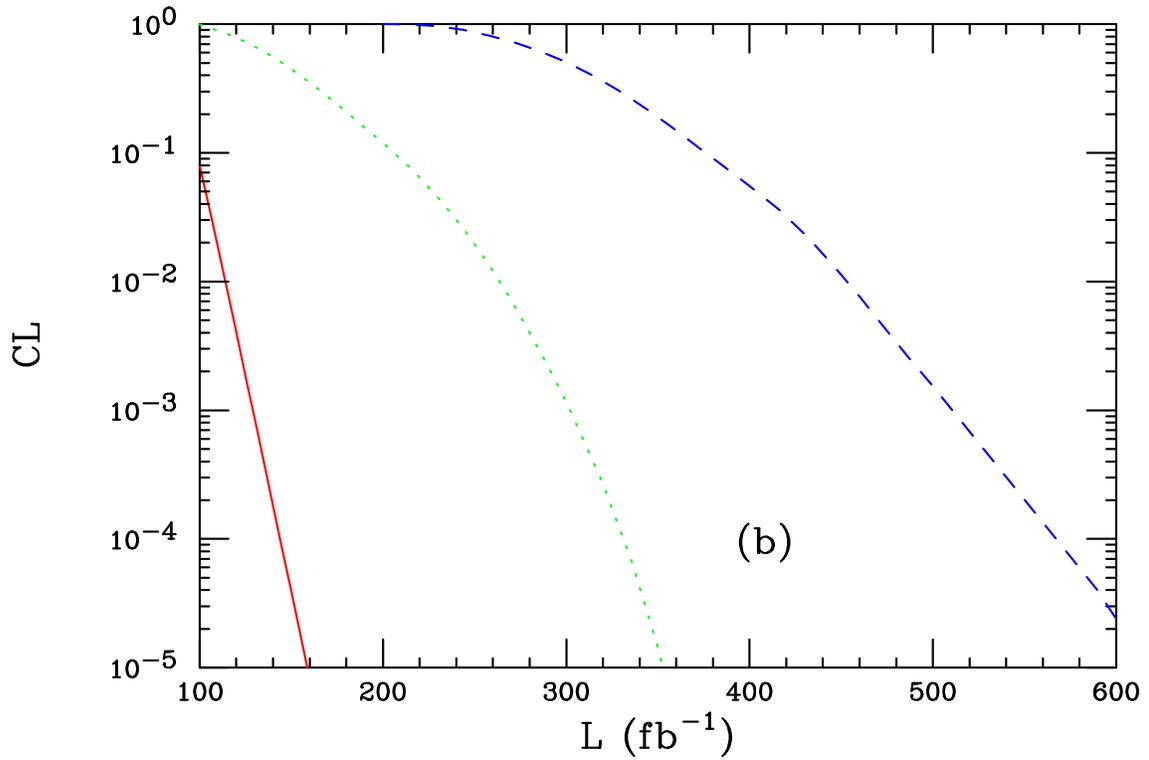,height=10.0cm,width=15cm,angle=90}}
\vspace*{0.0cm}
\caption{CL as a function of the integrated luminosity 
resulting from the coupling fits following from the analysis 
discussed in the text for both (a) a 500 GeV or a (b) 1 TeV $e^+e^-$ collider. 
In (a) the solid(dash-dotted,dotted) curve corresponds to a first KK 
excitation mass of 4(5,6) TeV. In (b) the solid(dotted,dashed) curve 
corresponds to a first KK mass of 5(6,7) TeV.}
\end{figure}
\vspace*{0.4mm}

The only problem with the fit shown in the figure is that 
the $\chi^2$ is very large leading to a very small confidence level, \ie, 
$\chi^2/d.o.f=95.06/58$ or P=CL=$1.55\times 10^{-3}$! (We note that this result 
is not very sensitive to the assumption of $90\%$ beam polarization; $70\%$ 
polarization leads to almost identical results.) For an ordinary $Z'$ it has 
been shown that fits of much higher quality, based on confidence level values, 
are obtained by this same procedure. 
Increasing the integrated luminosity can be seen to 
only make matters worse. Fig.5 shows 
the results for the CL following the above approach as we vary both the 
luminosity and the mass of the first KK excitation at both 500 GeV and 1 TeV 
$e^+e^-$ linear colliders. From this figure we see that the resulting CL 
is below $\simeq 10^{-3}$ for a first KK excitation with a mass of 4(5,6) 
TeV when the 
integrated luminosity at the 500 GeV collider is 200(500,900)$fb^{-1}$ whereas 
at a 1 TeV  for excitation masses of 5(6,7) TeV we require luminosities of 
150(300,500)$fb^{-1}$ to realize this same CL. Barring some unknown systematic 
effect the only conclusion that one could draw from such bad fits is that the 
hypothesis of a single $Z'$, and the existence of no other new physics, 
is simply {\it wrong}.  
If no other exotic states are observed below the first KK mass at the LHC this 
result would give very strong indirect evidence that something more unusual 
that a conventional $Z'$ had been found. The problem from the experimental 
point of view would be to wonder what fitting hypothesis to try next as there 
are so many possibilities to try. For example, one can imagine trying a two 
$Z'$ scenario with the first at 4 TeV, as discovered by the LHC, and with the 
second $Z'$ at 6 or more TeV, beyond the range of the LHC. 
Eventually one might try repeating the above fitting 
procedure allowing for two essentially degenerate new gauge bosons with 
different leptonic couplings could then be shown to yield a good fit to the 
data. Furthermore, it is clear from the discussion that all of the analysis 
performed above will go through in an almost identical manner in the case of 
more than one extra dimension yielding qualitatively similar results.

\section{Lepton Colliders on (and near) the Resonance}

In order to be completely sure of the nature of the first KK excitation, we 
must produce it directly at a higher energy lepton collider and sit on and 
near the peak of the KK resonance. To reach this mass range will most likely 
require either CLIC technology{\cite {clic}} or a Muon Collider{\cite {mumu}}. 
Recall that the mass of the KK resonance is already quite well known from 
data from the LHC so that the center of mass energy of the Muon Collider can 
be chosen near this value. 

The first issue to address is the quality of the degeneracy 
of the $\gamma^{(1)}$ and $Z^{(1)}$ states. If any part of the Higgs boson(s) 
whose 
vacuum expectation(s) value breaks the SM down to $U(1)$ is on the `wall' then 
the SM $Z$ will mix with the $Z^{(n)}$ slightly shifting all their masses; 
due to the remaining $U(1)$ gauge invariance this will not happen to the 
$\gamma^{(n)}$ thus implying a slight difference between the $\gamma^{(1)}$ 
and $Z^{(1)}$ masses. Based on the analyses in Ref.{\cite {fits}} we can get 
an idea of the maximum possible size of this fractional mass shift and we 
find it to be 
of order $\sim M_Z^4/M_{Z^{(1)}}^4$, an infinitesimal quantity for KK masses 
in the several TeV range. Thus even when mixing is included we find that the 
$\gamma^{(1)}$ and $Z^{(1)}$ states remain very highly degenerate so that even 
detailed lineshape measurements may not be able to 
distinguish the $\gamma^{(1)}/Z^{(1)}$ composite state from that of a $Z'$. 
We thus must turn to other parameters in order to separate these two cases.

\vspace*{-0.8cm}
\nn
\begin{figure}[htbp]
\centerline{
\psfig{figure=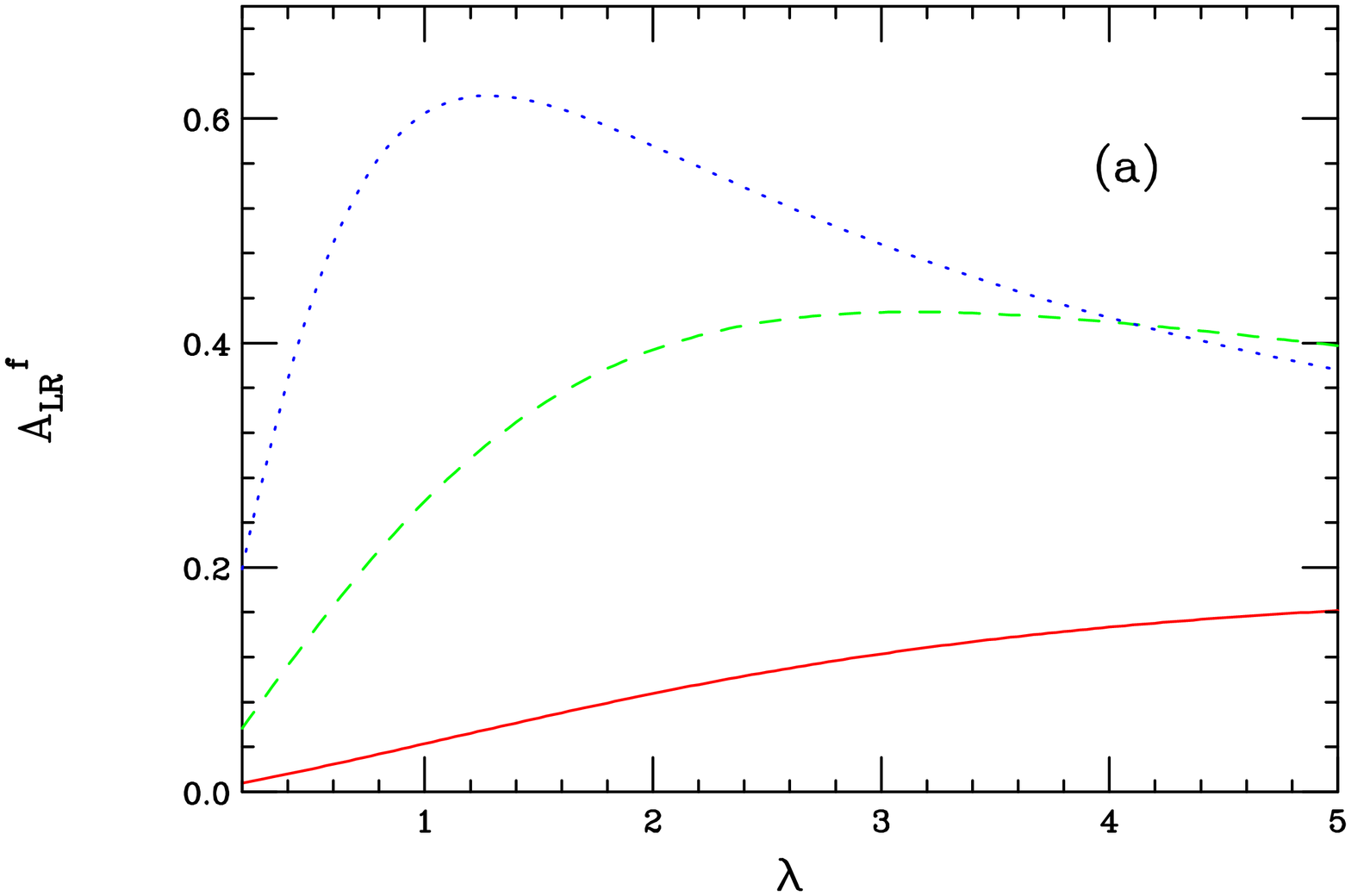,height=10.0cm,width=15cm,angle=90}}
\vspace*{9mm}
\centerline{
\psfig{figure=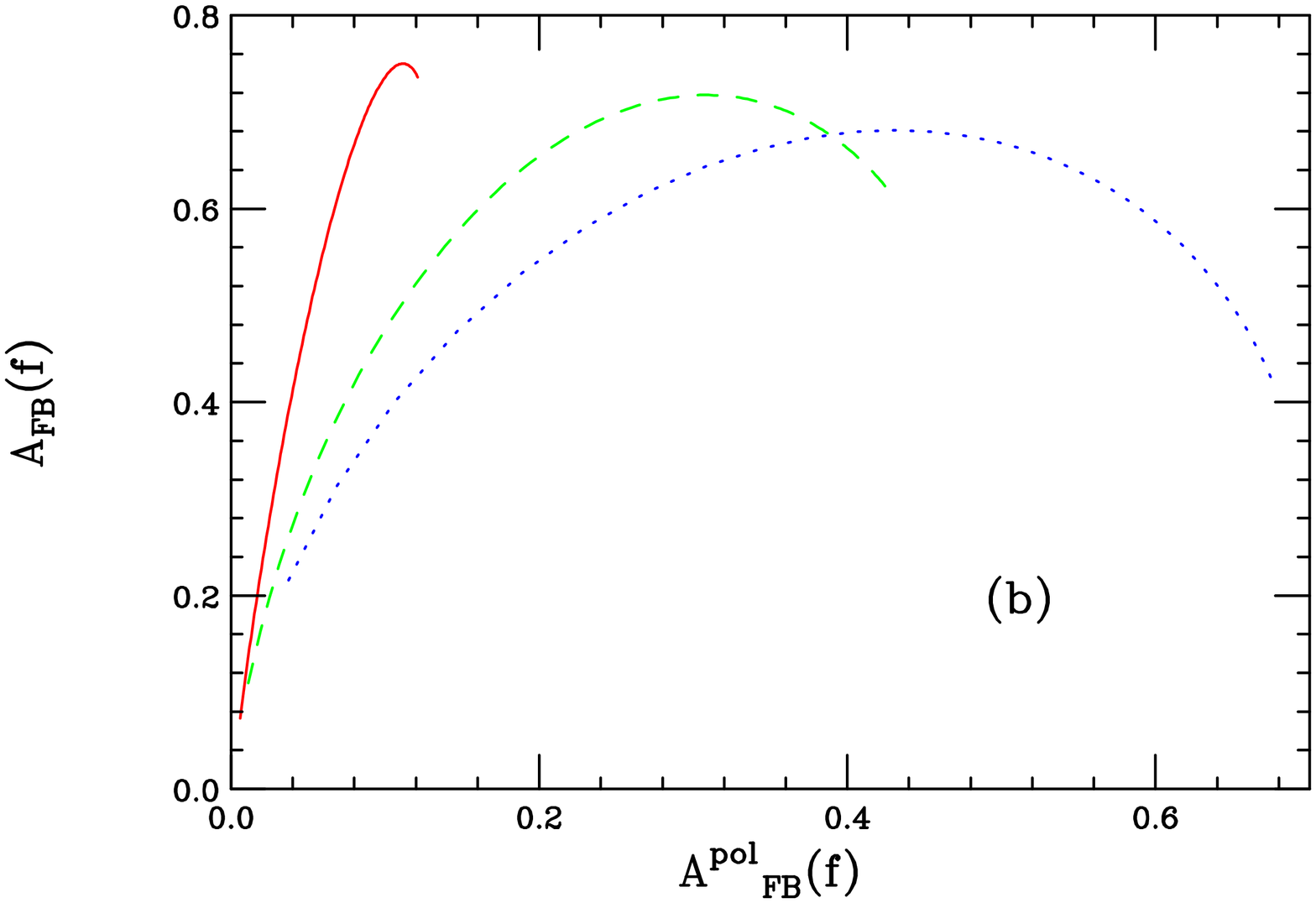,height=10.0cm,width=15cm,angle=90}}
\vspace*{0.0cm}
\caption{(a) $A_{LR}^f$ as a function of the parameter $\lambda$ for 
$f=\ell$(solid), $f=c$(dashed) and $f=b$(dots). (b) Correlations between 
on-peak observables for the same three cases as shown in (a). $\lambda$ 
varies from 0.2 to 5 along each curve.}
\end{figure}
\vspace*{0.4mm}

Sitting on the resonance there are a very large number of quantities that can 
be measured: the mass and apparent 
total width, the peak cross section, various partial 
widths and asymmetries \etc. From the $Z$-pole studies at SLC and LEP, we 
recall a few important tree-level results which we would expect to apply 
here as well 
provided our resonance is a simple $Z'$. First, we know that the value of 
$A_{LR}=[A_e=2v_ea_e/(v_e^2+a_e^2)]$, as measured on the $Z$ by SLD, does not 
depend on the fermion flavor of the final state and second, that the 
relationship $A_{LR}\cdot A_{FB}^{pol}(f)=A_{FB}^f$ holds, where 
$A_{FB}^{pol}(f)$ is the polarized Forward-Backward asymmetry as measured for 
the $Z$ at SLC and $A_{FB}^f$ is the usual Forward-Backward asymmetry. The 
above relation is seen to be trivially satisfied on the $Z$(or on a $Z'$) since 
$A_{FB}^{pol}(f)={3\over 4}A_f$ and $A_{FB}^f={3\over 4}A_eA_f$. Both of these 
relations are easily shown to fail in the present case of a `dual' resonance 
though they will hold if only one particle is resonating. 

\vspace*{-0.8cm}
\nn
\begin{figure}[htbp]
\centerline{
\psfig{figure=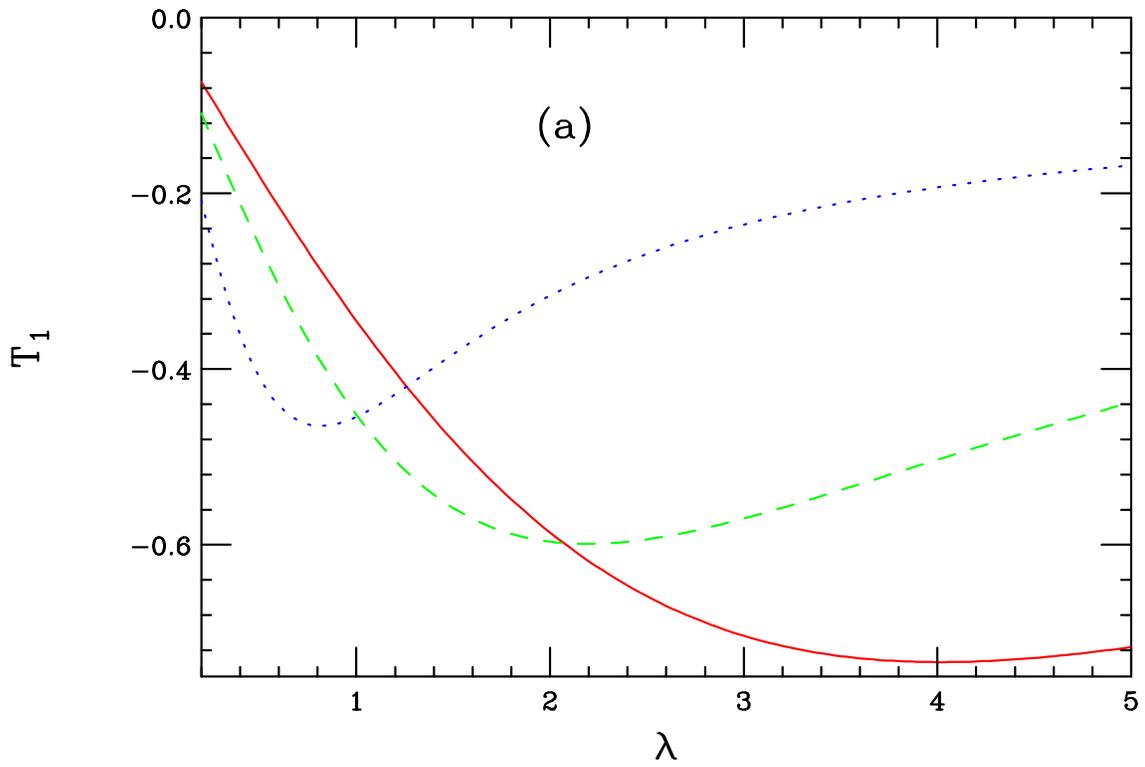,height=10.0cm,width=15cm,angle=90}}
\vspace*{9mm}
\centerline{
\psfig{figure=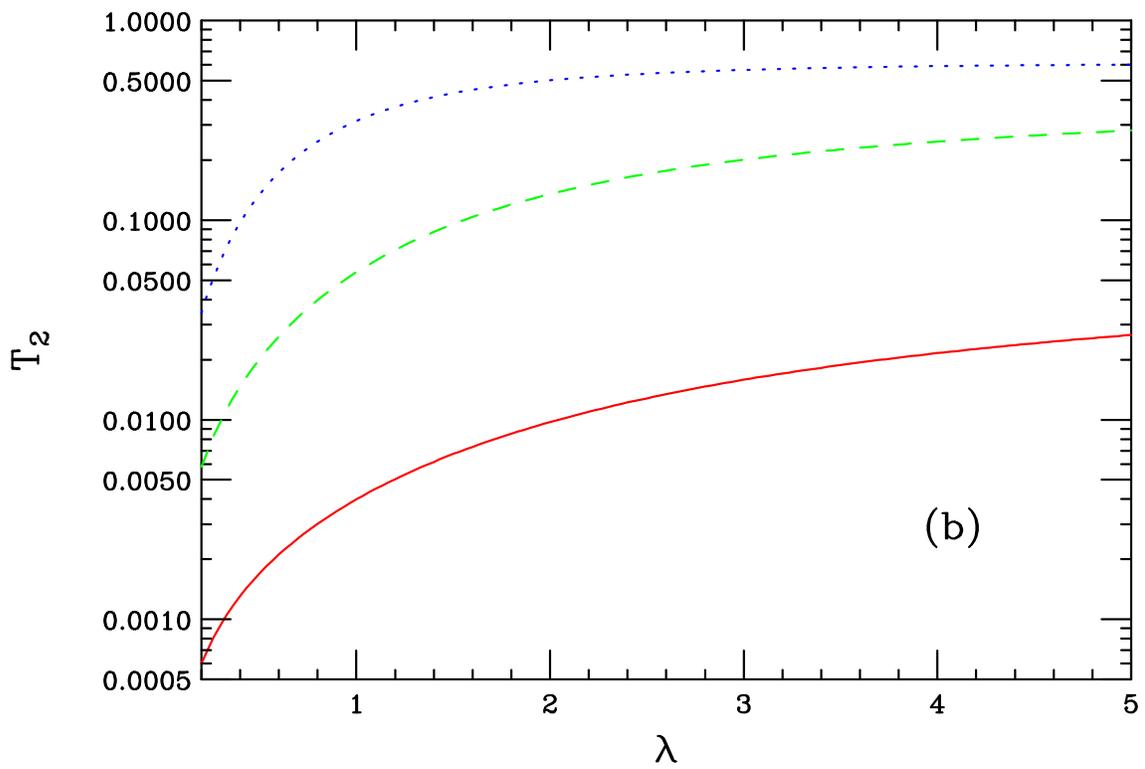,height=10.0cm,width=15cm,angle=90}}
\vspace*{0.0cm}
\caption{The quantities (a)$T_1$ and (b)$T_2$ as functions of the parameter 
$\lambda$. The curves are labelled as in the previous figure.}
\end{figure}
\vspace*{0.4mm}

A short exercise 
shows that in terms of the couplings to $\gamma^{(1)}$, which we will call 
$v_1,a_1$, and $Z^{(1)}$, now called $v_2,a_2$, these same observables can 
be written as 
\begin{eqnarray}
A_{FB}^f &=& {3\over 4} {A_1\over D}\nonumber \\
A_{FB}^{pol}(f) &=& {3\over 4} {A_2\over D}\nonumber \\
A_{LR}^f &=& {A_3\over D}\,,
\end{eqnarray}
where $f$ labels the final state fermion and we have defined the coupling 
combinations
\begin{eqnarray}
D &=& (v_1^2+a_1^2)_e(v_1^2+a_1^2)_f+R^2(v_2^2+a_2^2)_e(v_2^2+a_2^2)_f+
2R(v_1v_2+a_1a_2)_e(v_1v_2+a_1a_2)_f\nonumber \\
A_1 &=& (2v_1a_1)_e(2v_1a_1)_f+R^2(2v_2a_2)_e(2v_2a_2)_f+2R(v_1a_2+v_2a_1)_e
(v_1a_2+v_2a_1)_e\nonumber \\
A_2 &=& (2v_1a_1)_f(v_1^2+a_1^2)_e+R^2(2v_2a_2)_f(v_2^2+a_2^2)_e+2R(v_1a_2
+v_2a_1)_f(v_1v_2+a_1a_2)_e\nonumber \\
A_3 &=& (2v_1a_1)_e(v_1^2+a_1^2)_f+R^2(2v_2a_2)_e(v_2^2+a_2^2)_f+2R(v_1a_2
+v_2a_1)_e(v_1v_2+a_1a_2)_f\,,
\end{eqnarray}
with $R$ is the ratio of widths $R=\Gamma_1/\Gamma_2$ and the 
$v_{1,2i},a_{1,2i}$ are 
the appropriate couplings for electrons and fermions $f$. Note that when $R$ 
gets either very large or very small we recover the usual `single resonance' 
results. 
Examining these equations we immediately note that $A_{LR}^f$ is 
now {\it flavor dependent} and that the relationship between observables is 
no longer satisfied:
\begin{equation}
A_{LR}^f\cdot A_{FB}^{pol}(f)\neq A_{FB}^f\,,
\end{equation}
which clearly tells us that we are actually producing more than one resonance. 
Note that the numerical values of all of these asymmetries, being only 
proportional to ratios of various couplings, are independent of how much 
damping the KK couplings experience due to the stiffness of the wall. In what 
follows we will limit ourselves to electroweak observables whose values are 
independent of the overall normalization of the couplings and the potential 
exotic decay modes of the first KK excitation.

Of course we need to verify that these single resonance 
relations are numerically badly broken 
before clear experimental signals for more than one resonance can be claimed. 
Statistics will not be a problem with any reasonable integrated luminosity 
since we are sitting on a resonance peak and certainly millions of events 
will be collected. Assuming decays to only SM final states we estimate that 
with an integrated luminosity of 100 $fb^{-1}$ a sample of approximately 
half a million lepton pairs will be obtainable for each flavor.
In principle, to be as model independent as possible in a numerical analysis, 
we should allow the widths $\Gamma_i$ to be greater than or equal to their SM 
values as such heavy KK states may decay to SM SUSY partners as well as to 
presently unknown exotic states. Since the expressions above only depend upon 
the ratio of widths, we let $R=\lambda R_0$ where $R_0$ is the value 
obtained assuming that the KK states have only SM decay 
modes. We then treat $\lambda$ as a free parameter in what follows 
and explore the range 
$1/5 \leq \lambda \leq 5$. Note that as we take $\lambda \to 0(\infty)$ we 
recover the limit corresponding to just a $\gamma^{(1)}(Z^{(1)})$ being 
present.

\vspace*{-0.8cm}
\nn
\begin{figure}[htbp]
\centerline{
\psfig{figure=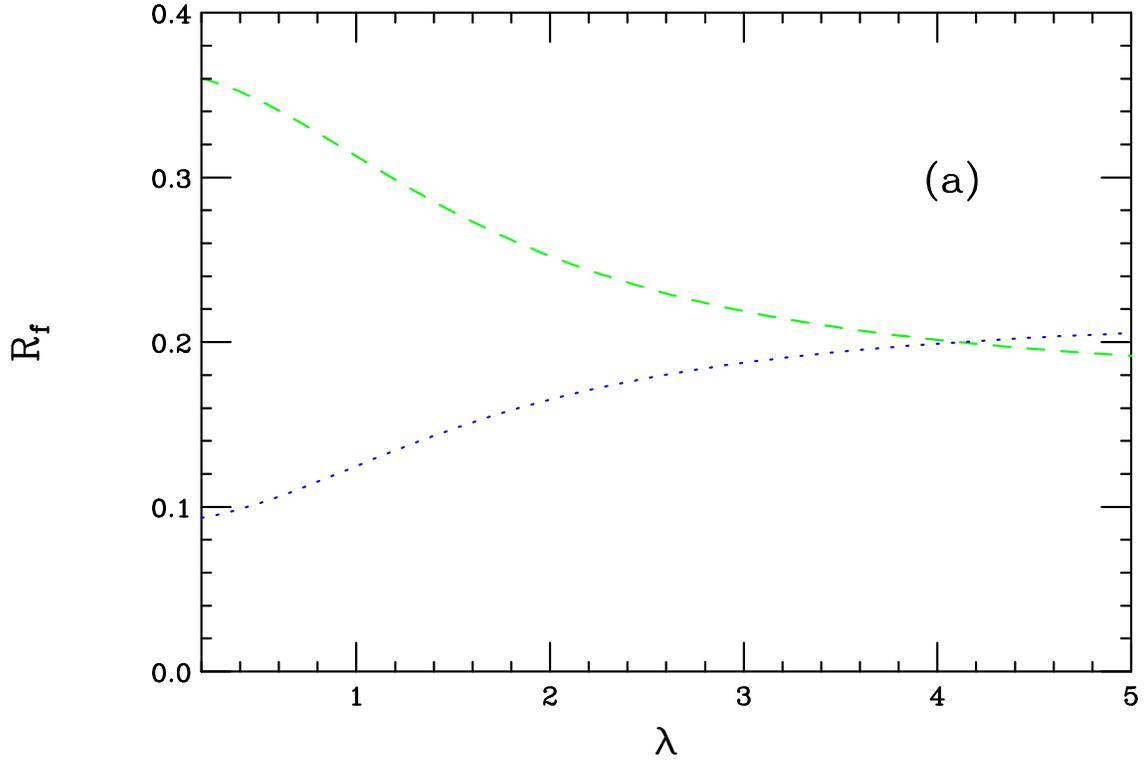,height=10.0cm,width=15cm,angle=90}}
\vspace*{9mm}
\centerline{
\psfig{figure=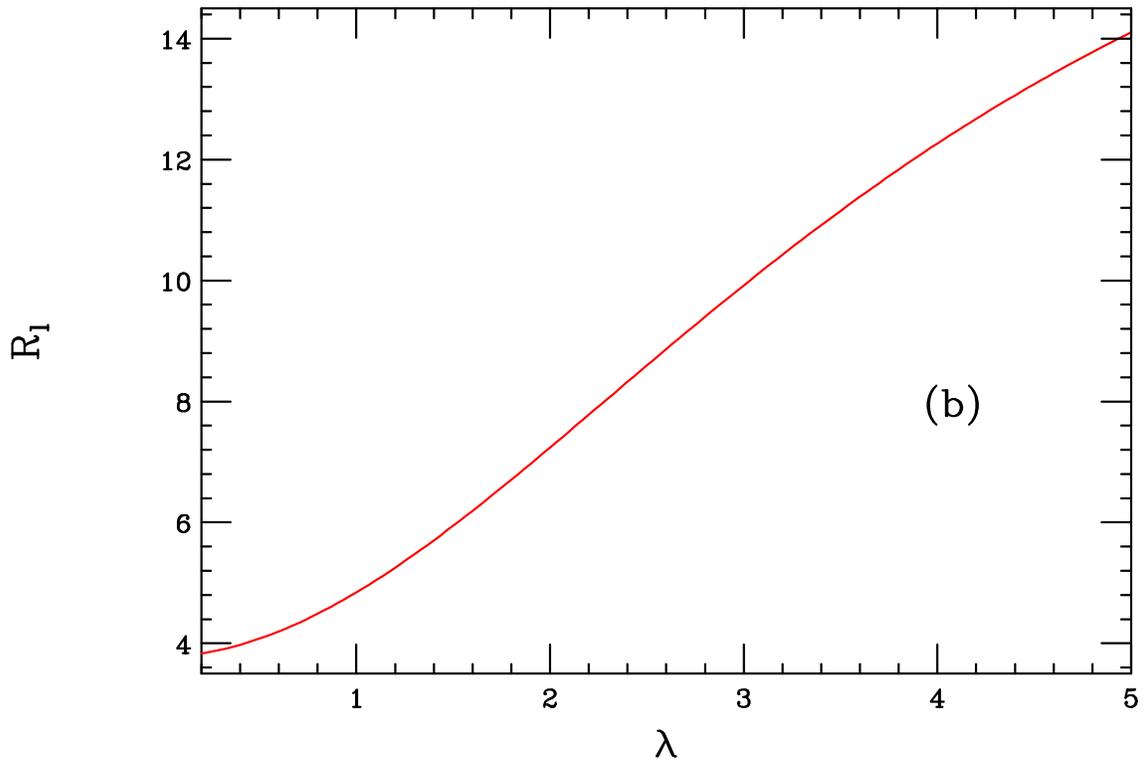,height=10.0cm,width=15cm,angle=90}}
\vspace*{0.0cm}
\caption{(a) The ratios of the $\bar b b$(dots) and $\bar c c$ cross sections 
to that for all hadrons on top of the 4 TeV KK excitation. (b) The ratio of 
the hadron cross section to that for leptons on the peak of the KK excitation.}
\end{figure}
\vspace*{0.4mm}

In Fig.6 we display the flavor dependence of $A_{LR}^f$ as a functions of 
$\lambda$. Note that as $\lambda \to 0$ the asymmetries vanish since the 
$\gamma^{(1)}$ has only vector-like couplings. In the opposite limit, for 
extremely large $\lambda$, the $Z^{(1)}$ couplings dominate and a common 
value of $A_{LR}$ will be obtained. It is quite clear, however, 
that over the range of reasonable values of $\lambda$, $A_{LR}^f$ is quite 
obviously flavor dependent. We also show in Fig.6 the correlations between the 
observables $A_{FB}^{pol}(f)$ and $A_{FB}(f)$ which would be flavor 
independent if only a single resonance were present. From the figure we see 
that this is clearly not the case. Note that although $\lambda$ is an 
{\it a priori} unknown parameter, once any one of the electroweak observables 
are measured the value of $\lambda$ will be directly determined. Once $\lambda$ 
is fixed, then the values of all of the other asymmetries, as well as the 
ratios of various partial decay widths, are all completely fixed for the KK 
resonance with uniquely predicted values. 

In order to further numerically probe the breakdown of the relationship between 
the on-resonance 
electroweak observables in Eq. (5) we consider two related quantities: 
\begin{eqnarray}
T_1 &=& A_{LR}^f\cdot A_{FB}^{pol}(f)-A_{FB}^f\nonumber \\
T_2 &=& {A_{LR}^f\cdot A_{FB}^{pol}(f) \over {A_{FB}^f}}\,,
\end{eqnarray}
the first of which should vanish while the second should be equal to unity 
independently of the fermion flavor if only a single 
resonance were present. In this case, both these variables have values far 
away from these expectations and are shown in Fig.7 as functions of $\lambda$. 

For completeness, we show in Fig.8 the ratio of partial widths 
$R_{b,c}=\Gamma_{b,c}/\Gamma_{had}$ and $R_{\ell}=\Gamma_{had}/\Gamma_{\ell}$. 
Once the parameter $\lambda$ is fixed the corresponding values of these 
observables become uniquely determined. The values of these quantities will 
help to pin down the nature of this resonance as a combined 
$Z^{(1)}/\gamma^{(1)}$ KK excitation. Recall that once $\lambda$ 
is fixed, then the values of all of the other asymmetries, as well as the 
ratios of various partial decay widths, are all completely fixed for the KK 
resonance with uniquely predicted values. These can then be compared to the 
data extracted by the Muon Collider and will demonstrate that the double 
resonance is composed of an excited photon and $Z$, the hallmark of the KK 
scenario. 

In Figs.9 and 10 we show that although on-resonance measurements of the 
electroweak 
observables, being quadratic in the $Z^{(1)}$ and $\gamma^{(1)}$ couplings, 
will not distinguish between the conventional KK scenario and that of the 
AS, the data below the peak in the hadronic channel 
will easily allow such a separation. The cross section and asymmetries for 
$e^+e^-\to \mu^+\mu^-$ (or vice versa) is, of course, the same in both cases. 
The combination 
of on and near resonance measurements will thus completely determine the 
nature of the resonance. Off-resonance measurements can be made possible 
through the use of `radiative returns' wherein the emission of initial state 
photons allow one to perform as scan at all energies below the collider 
design $\sqrt s$.

We note that all of the above analysis will go through essentially unchanged 
in any qualitative way 
when we consider the case of the first KK excitation in a theory with more 
than one extra dimension.

\vspace*{-0.8cm}
\nn
\begin{figure}[htbp]
\centerline{
\psfig{figure=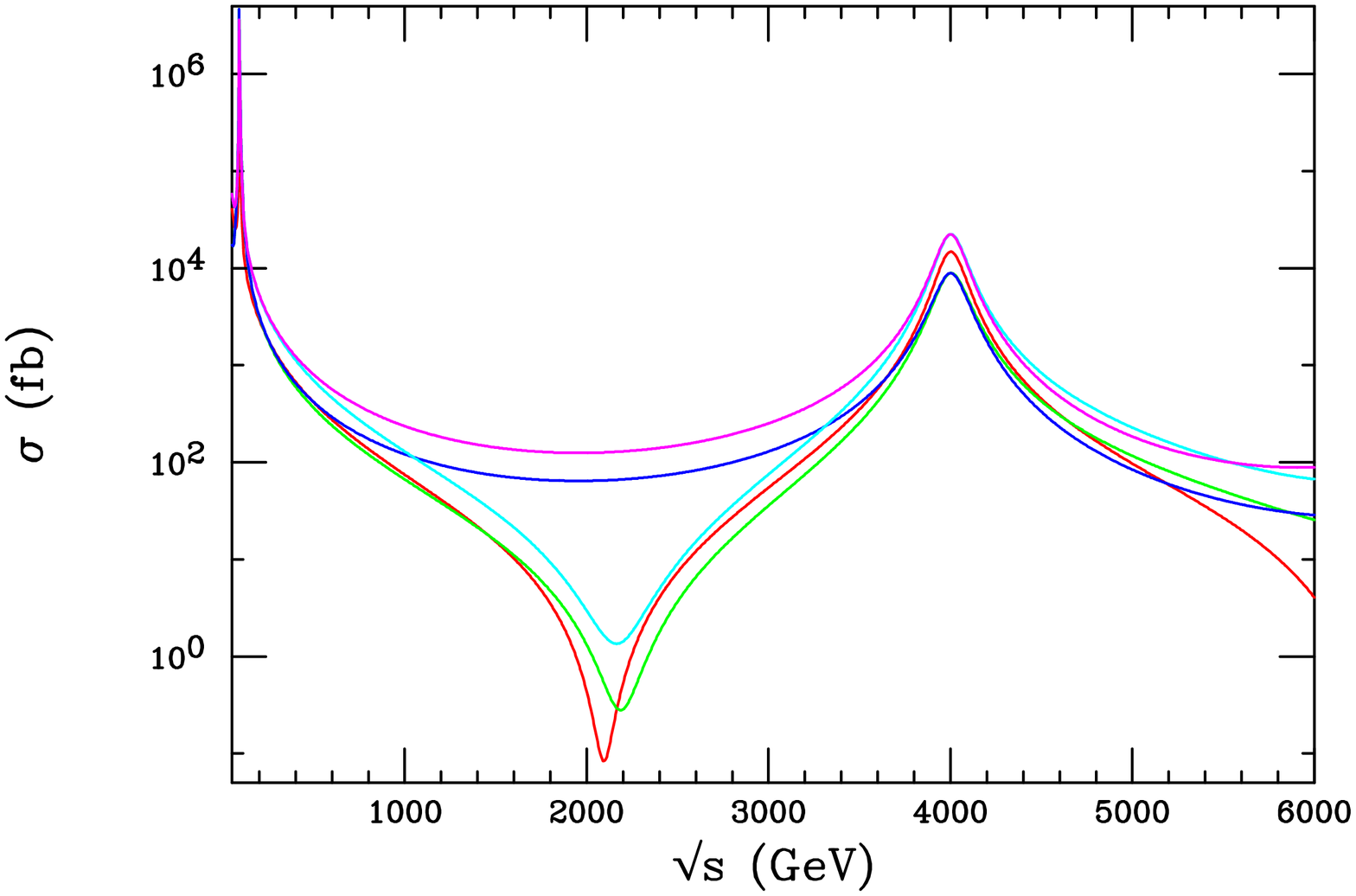,height=10.0cm,width=15cm,angle=90}}
\vspace*{9mm}
\centerline{
\psfig{figure=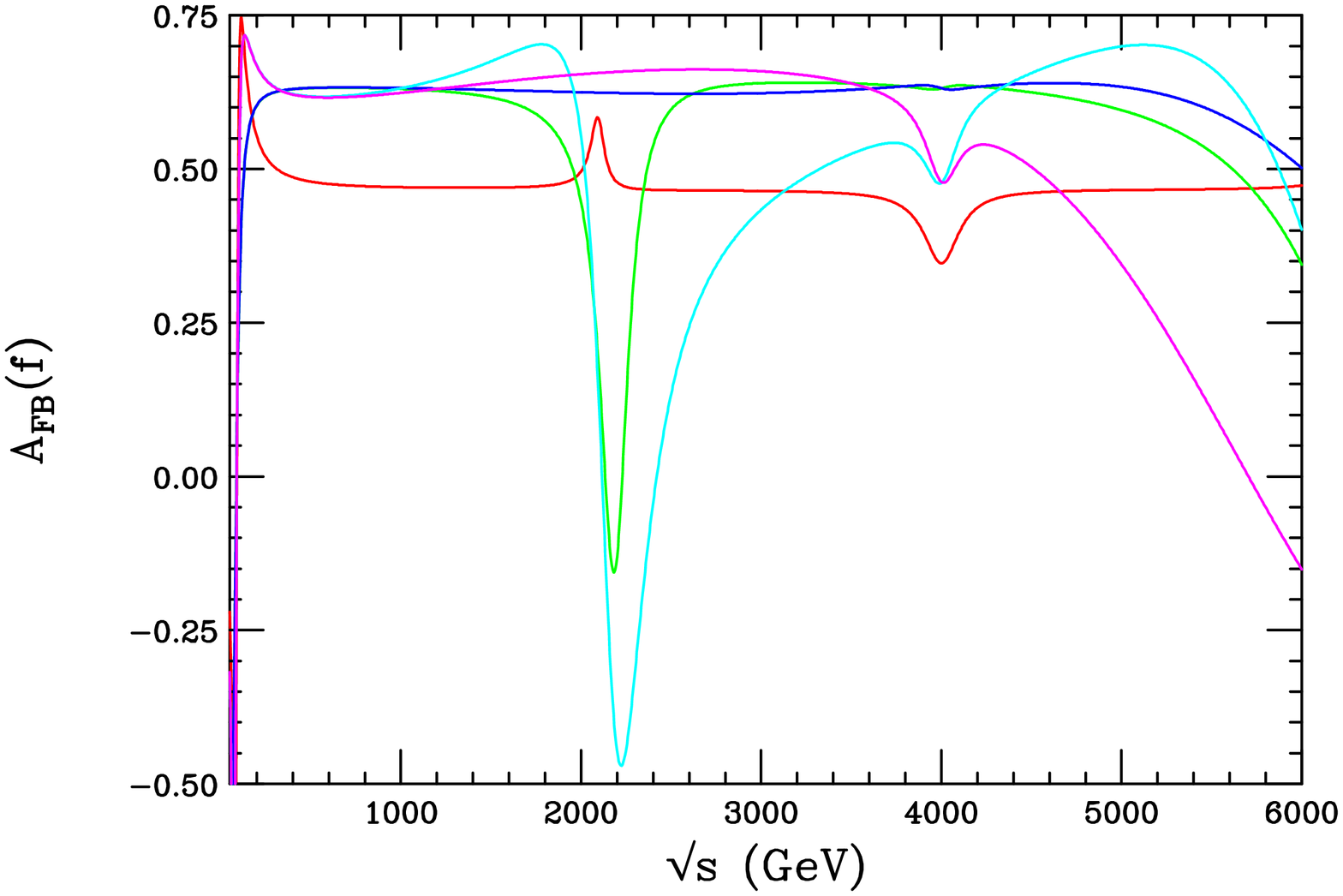,height=10.0cm,width=15cm,angle=90}}
\vspace*{0.0cm}
\caption[*]{Cross sections and $A_{FB}$ for $e^+e^-\to \mu^+\mu^-$
$b\bar b$ and $c\bar c$ at high energy lepton colliders in both the 
`conventional' scenario and that of AS{\cite {schm}}. The red curve 
applies for the $\mu$ final state in either model whereas the green(blue) and 
cyan(magenta) curves label the $b$ and $c$ final states for the 
`conventional'(AS) scenario.}
\end{figure}
\vspace*{0.4mm}
\vspace*{-0.8cm}
\nn
\begin{figure}[htbp]
\centerline{
\psfig{figure=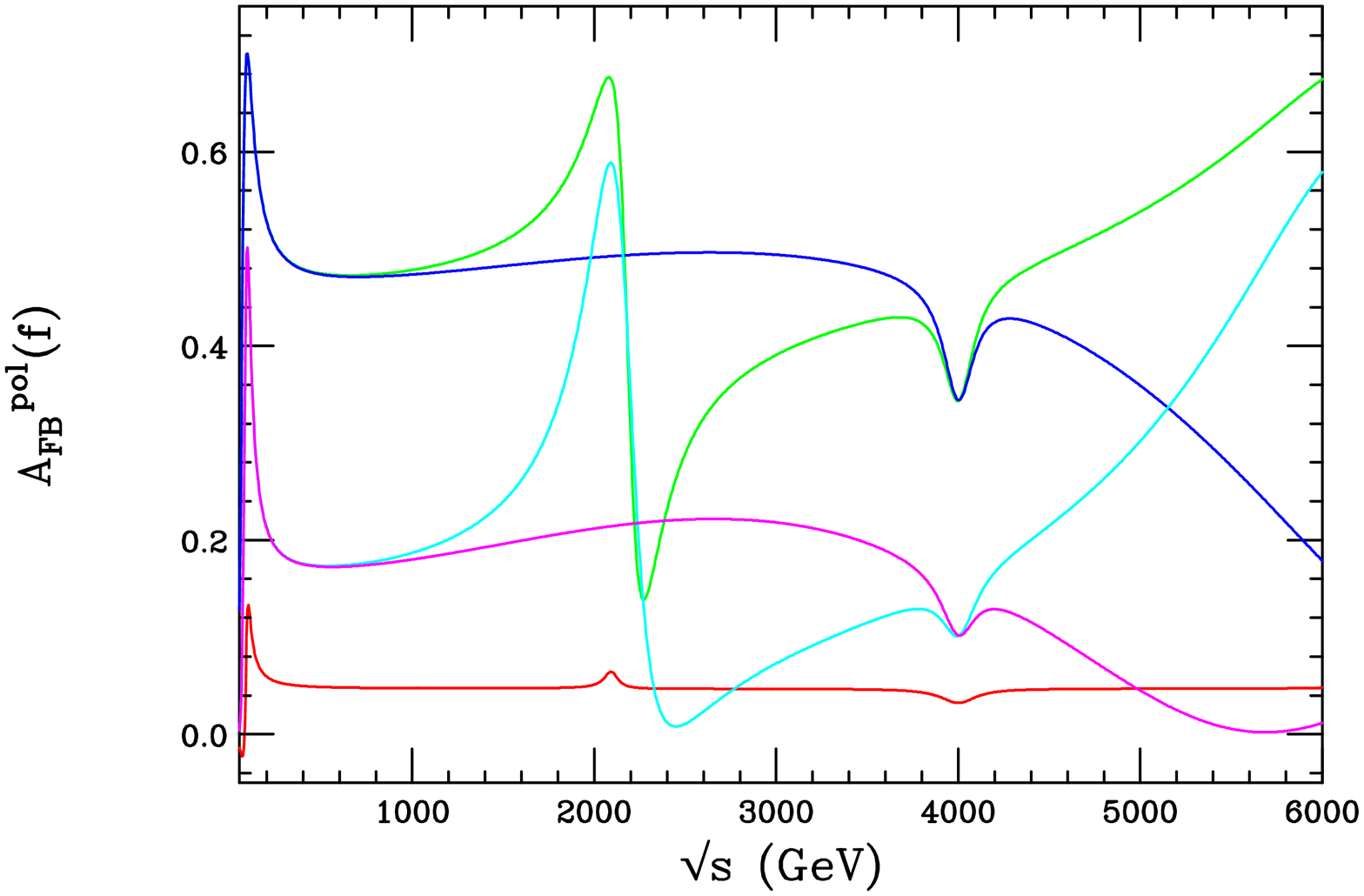,height=10.0cm,width=15cm,angle=90}}
\vspace*{9mm}
\centerline{
\psfig{figure=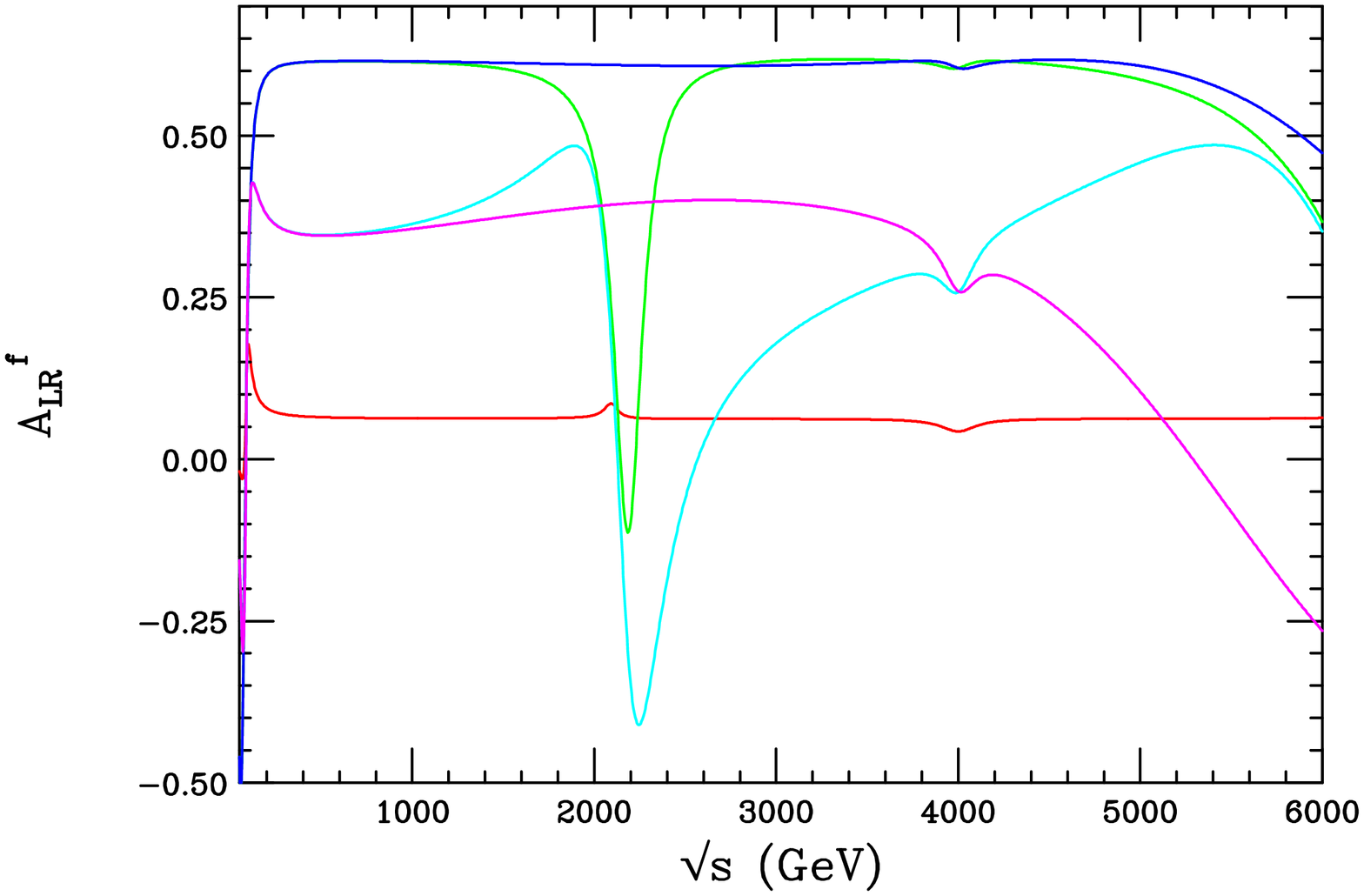,height=10.0cm,width=15cm,angle=90}}
\vspace*{0.0cm}
\caption[*]{Same as the previous figure but now for the observables 
$A_{LR}^f$ and $A_{FB}^{pol}(f)$.}
\end{figure}
\vspace*{0.4mm}

\section{Summary and Conclusions}

Present data strongly indicates that if KK excitations of the SM gauge bosons 
exist then the masses of the first tower excitations probably lie near or 
above 4 TeV if there is only one extra dimension and the KK states have 
naive couplings to the SM fermions. For KK masses in the 4-6 TeV 
range, the LHC will have sufficient 
luminosity to allow detection of these new particles although, for the case 
of the gluon excitation, any resonance-like structure might be too smeared out 
to be observable as a peak. Unfortunately, however, the second set of SM 
excitations will be far too heavy to be produced and thus the exciting 
`recurring resonance' signature one anticipates from KK theories is lost. 
In addition, 
since the first excitations are so massive, their detailed properties are not 
discernable so that the KK excitations of the $Z$ and $\gamma$ will appear to 
be a single $Z'$ peak. Thus as far as experiment at the LHC can determine 
we are left with what appears to be a degenerate 
$Z'/W'$ pair, something which can occur in many extended electroweak models. 
In this paper we have sought to resolve this problem based on information that 
can be gathered at lepton colliders in operation both far below and in the 
neighborhood of the first KK excitation mass. To perform this task and to 
solidify the above arguments we have taken the following steps obtaining 
important results along the way:

\begin{itemize}

\item  All constraints on the common mass, $M_1$, of the first SM KK 
excitations, apart from direct collider searches, rely on a two step process 
involving a parameter (or set of parameters) such as $V$ introduced in Eq.(1). 
First, given a set of data and the assumption that multiple new physics 
sources are not conspiring to distort the result, a bound on $V$ is obtained. 
The problem lies in the second step, \ie, converting the bound on $V$ into 
one on $M_1$. This is usually done in the case of one extra dimension where one 
naively sums over the entire tower of KK modes which converges numerically. 
However, we know that this convergence property no longer holds when we 
consider the case of more than one extra dimension which makes $V$ difficult to 
interpret in a more general context. Furthermore, the exact nature of the sum 
depends on the details of the compactification scenario when we extend the 
analysis to the case of more than one dimension. As specific examples we 
explored the case of two extra dimensions assuming either a 
$S^1/Z_2\times S^1/Z_2$, $Z_3$ or $Z_6$ 
compactification employing both a direct cut off of the tower sum as well as 
the better motivated exponential damping of the higher mode couplings. 
Depending upon the details of this cut off procedure in the case of more than 
one extra dimension we have shown that the lower limit on the mass of the 
first KK excitation arising from the bound on $V$ may lie outside of the range 
accessible to the LHC unless the parameter controlling the cut off, $n_{max}$, 
is quite small. Neither of these cut off approaches significantly 
alter our conclusions for the case of only one extra dimension. 

\item  Given the mass of the apparent $Z'$ resonance from the LHC, a low energy 
lepton collider can be used to attempt an extraction of it's leptonic 
couplings from a simultaneous fit to a number of distinct observables. While 
we demonstrated that a reasonably small region of 
the allowed parameter space will be selected by the fit, the confidence level 
was found to be very small in contrast to what happens in the case of an 
ordinary $Z'$. For example, a 500 GeV collider with an integrated luminosity 
of $\simeq 200$ $fb^{-1}$ probing a KK excitation with a mass of 4 TeV would 
obtain a fit probability of only $\simeq 10^{-3}$. This analysis was then 
generalized for both other KK masses and for a 1 TeV collider. We then argued 
that such an analysis would strongly indicate that the apparent $Z'$ observed 
by the LHC is {\it not} a single resonance. However, 
a fit allowing for the existence 
of a degenerate double resonance would yield an acceptable $\chi^2$. 

\item  Employing a lepton collider with sufficient center of mass energy, data 
can be taken at or near the first KK resonance. In this case we demonstrated 
that the familiar relationships between electroweak observables, in particular 
the various on-pole asymmetries, no longer hold in the presence of two 
degenerate resonances. For example, $A_{LR}$ now becomes flavor dependent and 
the factorization relationship $A_{LR}\cdot A_{FB}^{pol}(f)= A_{FB}^f$ was 
demonstrated to fail. The values of all of the on-pole observables was shown 
to be uniquely determined once the value of a single parameter, $\lambda$, 
which describes the relative total widths of the $\gamma^{(1)}$ and $Z^{(1)}$, 
is known. If both KK excitations only decay to SM particles, then $\lambda=1$. 
Furthermore, we showed that data taken off resonance can be used to distinguish 
among various models of the KK couplings and the localization of the fermions 
on the wall.

\end{itemize}

It is clear both the LHC and lepton colliders will be necessary to explore the 
physics of KK excitations.

\vskip1.0in

\noindent{\Large\bf Acknowledgements}

The author would like to thank J.L. Hewett, J. Wells, I. Antoniadis, P. Nath, 
N. Arkani-Hamed, H. Davoudiasl, M. Schmaltz, Y. Grossman, M. Masip and 
F. del Aguila for 
discussions related to this work. He would also like to thank the members of 
the CERN Theory Division, where this work was begun, for their hospitality.

\newpage

%
\def\MPL #1 #2 #3 {Mod. Phys. Lett. {\bf#1},\ #2 (#3)}
\def\NPB #1 #2 #3 {Nucl. Phys. {\bf#1},\ #2 (#3)}
\def\PLB #1 #2 #3 {Phys. Lett. {\bf#1},\ #2 (#3)}
\def\PR #1 #2 #3 {Phys. Rep. {\bf#1},\ #2 (#3)}
\def\PRD #1 #2 #3 {Phys. Rev. {\bf#1},\ #2 (#3)}
\def\PRL #1 #2 #3 {Phys. Rev. Lett. {\bf#1},\ #2 (#3)}
\def\RMP #1 #2 #3 {Rev. Mod. Phys. {\bf#1},\ #2 (#3)}
\def\NIM #1 #2 #3 {Nuc. Inst. Meth. {\bf#1},\ #2 (#3)}
\def\ZPC #1 #2 #3 {Z. Phys. {\bf#1},\ #2 (#3)}
\def\EJPC #1 #2 #3 {E. Phys. J. {\bf#1},\ #2 (#3)}
\def\IJMP #1 #2 #3 {Int. J. Mod. Phys. {\bf#1},\ #2 (#3)}

\end{document}